\documentclass[a4paper,11pt]{article}
\pdfoutput=1 % if your are submitting a pdflatex (i.e. if you have
\usepackage{jcappub} % for details on the use of the package, please
\usepackage[T1]{fontenc} % if needed
\usepackage{listings}

\usepackage{physics}
\usepackage{xcolor}
\usepackage{inconsolata}
\usepackage{float} % adds the [H] parameter for figures
\usepackage{bbm} % adds fancy 1 for indicator function
\lstnewenvironment{todo}[1][]
{
\lstset{
    basicstyle=\ttfamily,
    backgroundcolor=\color{yellow!20},
    #1
}
}{}
\newcommand{\vv}{{\rm{v}}}
\newcommand{\lr}[1]{\left({#1}\right)}

\newcommand{\photonunit}{$\text{photon}\, \text{cm}^{-2} \, \text{s}^{-1} \, \text{sr}^{-1} \, \text{\AA}^{-1}$}

\definecolor{editblue}{rgb}{0.12, 0.1, 0.9}
\newcommand{\edit}[1]{\textcolor{black}{#1}}

\newcommand{\editt}[1]{\textcolor{black}{#1}}

\usepackage{totcount}

\def\rmd{\mathrm{d}} % define the differential d
\title{\boldmath The Glow of Axion Quark Nugget Dark Matter: (III) The Mysteries of the Milky Way UV Background}

\author[a, b]{Michael Sekatchev,}
\author[a]{Xunyu Liang,}
\author[a]{Fereshteh Majidi,}
\author[a]{Ben Scully,}
\author[a]{Ludovic Van Waerbeke,}
\author[a]{Ariel Zhitnitsky}
\affiliation[a]{Department of Physics and Astronomy,\\
University of British Columbia, \\
V6T 1Z1 Vancouver, BC, Canada}
\affiliation[b]{Department of Nuclear Engineering, \\
University of California, Berkeley, \\
94720 Berkeley, CA, USA}

\emailAdd{michaelsekatchev@berkeley.edu}

\abstract{
Axion quark nuggets (AQNs) are hypothetical objects with nuclear density that would have formed during the quark-hadron transition and could make up most of the dark matter today. These objects have a mass greater than a few grams and are sub-micrometer in size. They would also help explain the matter-antimatter asymmetry and the similarity between visible and dark components of the universe, i.e. \( \Omega_{\text{DM}} \sim \Omega_{\text{visible}} \). These composite objects behave as cold dark matter, interacting with ordinary matter and producing pervasive electromagnetic radiation. This work aims to calculate the FUV electromagnetic signature in a \( 1 \, \text{kpc} \) region surrounding the solar system, resulting from the interaction between antimatter AQNs and baryons. To this end, we use the high-resolution hydrodynamic simulation of the Milky Way, FIRE-2 Latte suite, to select solar system-like regions. From the simulated gas and dark matter distributions in these regions, we calculate the FUV background radiation generated by the AQN model. We find that the results are consistent with the FUV excess recently confirmed by the Alice spectrograph aboard New Horizons, which corroborated the FUV excess initially discovered by GALEX a decade ago. We also discuss the potential cosmological implications of our work, which suggest the existence of a new source of FUV radiation in galaxies, linked to the interaction between dark matter and baryons.}

\begin{document}
\maketitle
\flushbottom

\pagebreak

%#################################################################|
%#################################################################|
\section{Introduction}\label{sec:introduction}
%#################################################################|
%#################################################################|

The Galaxy Evolution Explorer (GALEX), as fully described in \citep{Martin:2005-GALEX}, conducted observations of the diffuse Galactic far-ultraviolet (FUV) background using its FUV imager ($1350$--$1750$ \AA).
The majority of the detected FUV background could be attributed to dust-scattered starlight. However, a significant isotropic residual flux of 200 \photonunit (or “photon units”) was identified after removing all known astrophysical sources \citep{Hamden:2013-GALEX-analysis}. Subsequent studies \citep{Henry:2015-uv-excess-galex,Akshaya:2018-uv-excess,Akshaya:2019-uv-excess} were unable to attribute this excess to scattered FUV light but determined that it must have a Galactic and astrophysical origin, giving rise to the so-called “mystery” of the cosmic diffuse UV background excess. The brightness of the GALEX FUV background was in close agreement with previous observations made by spacecraft located much farther from Earth (Dynamics Explorer, \citep{Fix:1989-Dynamics-Explorer}). This ruled out the possibility of a solar-system or terrestrial origin. 
The interpretation of the observations remains an open question, essentially because of the following puzzling characteristics: 

\begin{enumerate}
\item The "uniformity puzzle": The diffuse radiation appears to be remarkably uniform across both hemispheres, as illustrated in figures 7--10 of \cite{Henry:2015-uv-excess-galex}. This stands in stark contrast to the pronounced non-uniformity in the distribution of UV-emitting stars.

\item The "Galactic longitude puzzle": The diffuse radiation exhibits almost no dependence on Galactic longitude. This is in stark contrast to the distribution of the brightest UV-emitting stars, which are predominantly confined to the longitude range $180^\circ$-$\,360^\circ$. This strong discrepancy suggests that the diffuse background radiation is unlikely to originate from dust-scattered starlight.

\item The "Galactic latitude puzzle": The diffuse radiation becomes brighter toward lower Galactic latitudes at all Galactic longitudes. This behaviour contrasts with conventional modelling \cite{Henry:2015-uv-excess-galex}, which predicts very low brightness at low Galactic latitudes. 
This observation also suggests that the recorded diffuse emission has a Galactic, rather than extragalactic, origin. Indeed, extragalactic light cannot exhibit a strong variation toward lower Galactic latitudes. For the same reasons, the observations also conclusively indicate that this emission is celestial rather than terrestrial in nature.
  
\item The "non-correlation puzzle": The conventional model of UV diffuse radiation suggests that it should be correlated with the 100 $\mu$m thermal emission, as both are assumed to be linked to dust and its distribution in the galaxy. However, this assumption fails dramatically, as the 100 $\mu$m thermal emission is highly asymmetric and explicitly correlates with the localization of UV-emitting stars, whereas the UV diffuse emission is highly uniform and shows no correlation with the dust distribution. See Figure 14 in \cite{Henry:2015-uv-excess-galex}. 

\end{enumerate}

A recent analysis \citep{Murthy:2025-uv-excess-new-horizons} of observations from the Alice UV spectrograph on the New Horizons spacecraft ($912$--$1000$ \AA~and $1400$--$1800$ \AA), located 57 AU from the Sun—well beyond the range of most foreground contamination and zodiacal light—confirmed the excess previously identified by GALEX. While approximately half of the observed intensity can be attributed to known sources, the remaining signal could not be explained by known UV sources.

The puzzling uniformity of the FUV excess \citep{Henry:2015-uv-excess-galex,Akshaya:2018-uv-excess} has also led to the consideration of dark matter annihilation as a potential source, however, all the reasonable candidates appear to be many orders of magnitude off: \edit{Namely, various possibilities have been proposed and discussed in \citep{Henry:2015-uv-excess-galex}, including the decay of a 27 eV $\tau$ neutrino, the annihilation of leading dark matter candidates such as Weakly Interacting Massive Particles (WIMPs) and QCD axions, and primordial black holes (PBHs). In the case of the $\tau$ neutrino, the expected background signal would consist of a single emission line from the decay photon, which is inconsistent with the broad, flat-shaped distribution observed in the GALEX data. For WIMPs, the signature is estimated to be at least 12 orders of magnitude lower than the GALEX excess. Based on this, it is hard to connect the excess to WIMP scattering in any form. By contrast, for axions, the reverse problem arises: the predicted flux is 12 orders of magnitude too bright and peaks in the X-ray band at 2 \AA~\citep{Henry:2015-uv-excess-galex}. As a result, conventional WIMPs and QCD axions are ruled out as viable sources of the excess. Finally, PBHs are shown in \citep{Henry:2015-uv-excess-galex} to predict a sharp emission peak in the gamma-ray spectrum (about 100 MeV) from Hawking evaporation, which is inconsistent with observation. While this peak could be shifted to the FUV band, it is difficult to naturally motivate this modification to the primordial mass function of the black holes.}
Interestingly, \citep{Henry:2015-uv-excess-galex} speculates that a composite dark matter particle could potentially produce the observed signal. Specifically, they wrote
"if the dark-matter particles should turn out to be composite particles, overall electrically neutral but involving electrically charged components (as in a neutron), then perhaps our second component could originate in collisions of those dark-matter particles with the nuclei of the interstellar medium".

This description precisely corresponds to the Axion Quark Nugget (AQN) dark matter model \citep{Zhitnitsky:2002-aqn-introduction}, in which dark matter consists of composite objects with masses typically \edit{$M_{\rm DM} \gtrsim 5~\rm g$}, nuclear density, and sub-micrometer size. This model is a generalization of Witten’s quark nugget hypothesis \citep{Witten:1984-nuggets, Farhi:1984-nuggets, deRujula:1984-nuggets}. Unlike conventional dark matter candidates such as WIMPs and axions (see review \citep{Marsh:2015-axion-cosmology} and references therein), AQNs interact strongly with baryonic matter. They form during the quark-hadron transition, at a time when matter and antimatter have not yet completely annihilated. Consequently, some AQNs exist in antimatter form, allowing them to collide with normal matter, leading to broadband electromagnetic radiation spanning from the radio to X-ray regime.

This work builds upon our recent studies, where the unique emission characteristics of AQNs were applied to large-scale structure \citep{Majidi:2024-aqn-glow-pt-1} and galaxy cluster analyses \citep{Sommer:2024-aqn-glow-pt-2}. In both cases, we found that the energy injection due to AQN annihilation with surrounding material not only remains consistent with existing constraints but also leads to specific predictions that could be tested by Euclid and JWST. The present study differs from these previous investigations in a crucial aspect: here, we explore the possibility that the AQN model could be the source of the observed FUV excess, aligning with the measured excess emission and potentially explaining the puzzling observations reported in \citep{Henry:2015-uv-excess-galex, Akshaya:2018-uv-excess, Akshaya:2019-uv-excess}, as initially proposed in \citep{Zhitnitsky:2021-uv-excess}. In this work, we take a significant step toward supporting this hypothesis by developing a comprehensive model of the FUV radiation generated by AQNs within the Milky Way and using hydrodynamical simulation to describe realistically the gas and dark matter distribution in a galactic environment similar to the one surrounding the solar system.

The organization of this paper is as follows. Section~\ref{sec:aqn} of this paper introduces the Axion Quark Nugget dark matter model, and discusses the annihilation mechanism responsible for producing a broadband Bremsstrahlung radiation. Section~\ref{sec:modeling-aqn} discusses the implementation details of the modelling of this annihilation interaction in a simulated Milky Way-like environment. The main results of this modelling are presented in section~\ref{sec:results}. Finally, section~\ref{sec:discussion} discusses these results, and concludes with presentation of future avenues for investigation.

%#################################################################|
%#################################################################|
\section{Axion Quark Nuggets}\label{sec:aqn}
%#################################################################|
%#################################################################|

This section provides a brief overview of the Axion Quark Nugget (AQN) model, with a particular emphasis on the AQN emission mechanism that serves as the foundation of this project. For a more detailed discussion on the formation and stability of AQNs, readers may refer to earlier papers in the The Glow of Axion Quark Nugget Dark Matter series \citep{Majidi:2024-aqn-glow-pt-1, Sommer:2024-aqn-glow-pt-2} or a recent review \citep{Zhitnitsky:2021-aqn-recent-review}.

\subsection{Characteristics of AQNs}

First introduced in \citep{Zhitnitsky:2002-aqn-introduction} as a potential explanation for the observed similarity between dark matter (DM) and visible matter (VM) densities, $\Omega_{\text{DM}}\sim\Omega_{\text{VM}}$, Axion Quark Nuggets are a proposed form of macroscopic dark matter. The model shares similarities with Witten's quark nuggets \citep{Witten:1984-nuggets}. AQNs are characterized as large, composite objects with nuclear density, possessing a mass \edit{$m_\text{AQN}\gtrsim 5\rm g$} and exhibiting a "cosmologically dark" cross-section to mass ratio, $\sigma/m_{\text{AQN}}\lesssim 10^{-10}~\rm cm^2g^{-1}$, consistent with the observational requirement for cold dark matter. The main distinguishing feature of the model is that AQNs can be composed of both {\it matter} and {\it antimatter} during the quarks-hadrons transition as a result of the charge segregation process (see, e.g., the brief review \citep{Zhitnitsky:2021-aqn-recent-review}). This charge segregation mechanism separates quarks from antiquarks due to the dynamics of the $\cal CP$-odd axion field. The resulting separation of baryon charges leads to the formation of quark nuggets and anti-quark nuggets at a similar (but not identical) rate, and the baryon net charge of the Universe remains zero. 

As mentioned, in the AQN scenario, $\Omega_{\rm DM}$ (representing here the matter and antimatter nuggets) and $\Omega_{\rm VM}$ naturally assume comparable densities, $\Omega_{\rm DM} \sim \Omega_{\rm VM}$. This arises because both components are proportional to the same fundamental dimensional parameter of the theory, $\Lambda_{\rm QCD}$. Consequently, the AQN model, by construction, simultaneously addresses two fundamental problems in cosmology: it provides an explanation for the baryon asymmetry of the Universe and accounts for the presence of dark matter with the appropriate density, $\Omega_{\rm DM} \sim \Omega_{\rm VM}$, without requiring any parameter fine-tuning.

AQNs have a nuclear density and a mass in the range of 5–500 g (see, e.g., Ref. \cite{Majidi:2024-aqn-glow-pt-1} and references therein). The geometrical radius of an AQN is related to its mass, $m_{\rm AQN}$, as follows: 

\begin{equation}
R \approx2.25\times10^{-5}{\rm\,cm}\left(\frac{m_{\rm AQN}}{16.7\rm,g}\right)^{1/3}. 
\end{equation} 

Since this work focuses on the electromagnetic emission of antimatter AQNs, we adopt the same convention as in \cite{Majidi:2024-aqn-glow-pt-1,Sommer:2024-aqn-glow-pt-2} and define the number density of AQNs as: 
\begin{equation} 
n_{\rm AQN} \equiv\frac{2}{3}\times\frac{3}{5}\times\frac{\rho_{\rm DM}}{m_{\rm AQN}c^2},
\end{equation} 

where the factor of $\frac{2}{3}$ accounts for the electromagnetic fraction of the AQN emission, the factor of $\frac{3}{5}$ represents the antimatter sector of the AQNs, $\rho_{\rm DM}$ is the dark matter energy density, and $c$ is the speed of light in vacuum. For simplicity, we assume that \edit{(i)} AQNs constitute the dominant component of dark matter and that \edit{(ii)} all AQNs have the same mass, $m_{\rm AQN}$. \edit{Assumption (i) is realistic and natural to assume because both AQNs and baryons have the same QCD origin. This is in contrast to other conventional dark matter candidates (e.g. WIMPs and axions), which are nonbaryonic and require fine tuning of model parameters. The detailed description can be found in review \cite{Zhitnitsky:2021-uv-excess} and original paper \cite{Ge:2018-aqn-formation-3}}. \edit{Assumption (ii) is reasonable to assume as a first approximation approach to the problem. As reviewed in \cite{Zhitnitsky:2021-uv-excess} and original paper \cite{Raza:2018gpb,Ge:2019-aqn-formation-4}, the mass distribution of AQN is a steep power function and can be approximated as a delta function. The evidence is based on the simulation of AQN formation and the observation of solar flares.}

It is specifically the antimatter nuggets that play a crucial role in the discussions that follow. These antimatter nuggets can collide with visible matter, leading to annihilation processes and subsequent radiation, as explained in section \ref{subsec:aqn-emission}. The analysis of the resulting electromagnetic emission in the UV frequency bands is the central focus of this work.

% \vp 

% \subsection{Formation and stability}\label{subsec:aqn-formation-stability}

% The AQN model relies on the existence of an axion field, with the AQNs forming during the quantum chromodynamic (QCD) phase transition in the early Universe from axion domain wall bubbles, further discussed in detail in \citep{Liang:2016-aqn-formation-1,Ge:2017-aqn-formation-2,Ge:2018-aqn-formation-3,Ge:2019-aqn-formation-4}. By accumulating quarks during the QCD transition, some bubbles in the axion field may persist, by forming an internal Fermi pressure to balance external pressure on the wall. As matter-antimatter annihilation continues in the plasma, so does the formation of matter and antimatter nuggets in axion domain wall bubbles, in the same proportions as the newly formed protons and neutrons outside. As cooling continues, all anti-baryons outside of the bubbles are depleted, but the AQNs persist in both matter and anti-matter form---the significant implication here is that, given that the Universe's total baryon charge is zero, the densities of DM and baryonic matter naturally align to the same order of magnitude. The quarks in the interior of the AQNs exist in a color superconducting phase, which at high enough pressures has a lower energy per baryon charge than in hadrons, showing that nuggets are more stable than protons and neutrons.

% \vp 
% \begin{todo}
% Likely add some more here - AQN mass discussion
% \end{todo}

\subsection{AQN emission mechanism}\label{subsec:aqn-emission}

In addition to the quarks contained within AQNs by the axion field, matter-based nuggets are surrounded by a cloud of electrons to maintain overall charge neutrality. Similarly, antimatter nuggets are enveloped by a cloud of positrons. It is this antimatter variant that is of particular interest: while the collision of matter AQNs with regular baryonic matter does not result in substantial energy exchange and thus produces negligible radiation, the interaction of antimatter AQNs with regular matter leads to annihilation events that release significant amounts of energy. Despite the strong interactions involved, the absence of an astrophysical prominent electromagnetic signature within the AQN framework can be attributed to the low number density of AQNs, which is a consequence of their large individual mass. As a result, the global rate of annihilation interactions remains relatively low.

This annihilation of antimatter AQNs with the surrounding regular matter generates a Bremsstrahlung emission signature spanning from radio to $\gamma$-rays. This radiation is produced by positrons in the electrosphere that have been heated by a fraction of the annihilation energy. The UV emission spectrum constitutes a component of this signal. The spectral emissivity $F_\nu$ from the surface of an AQN, expressed in units of $\rm[erg,cm^{-2}s^{-1}Hz^{-1}]$, is determined by the AQN temperature $T_{\rm AQN}$ \cite{Forbes:2008uf}:
\begin{equation}
\label{eqs:F_nu etc.}
\begin{aligned}
&F_\nu
=\frac{8\alpha^{5/2}}{45\hbar^2c^2}(k_B T_{\rm AQN})^3
\left(\frac{k_B T_{\rm AQN}}{m_ec^2}\right)^{\frac{1}{4}}
H\left(\frac{2\pi\hbar\nu}{k_BT_{\rm AQN}}\right)\,;  \\
&H(x)\approx\left\{
\begin{aligned}
&(1+x)e^{-x}\left(17-12\ln (x/2)\right)\,,
&{\rm if~} x<1\,; \\
&(1+x)e^{-x}(17+12\ln2)\,,
&{\rm if~}x\geq1\,,
\end{aligned}
\right.
\end{aligned}
\end{equation}
where $\nu$ is the frequency of the photon emission, $\alpha$ is the fine structure constant, $k_B$ is the Boltzmann constant, $\hbar$ is the reduced Planck constant, and $m_e$ is the electron mass.
% \footnote{An alternative emission mechanism was suggested in \cite{Flambaum:2021xub,Flambaum:2021awu,Flambaum:2022wcs}, but as we discuss in Appendix \ref{app:emission} many their claims are incorrect, including their estimates for the intensity and spectrum. Therefore, we continue with the original approach developed in \cite{Forbes:2008uf}.}

As shown in \cite{Majidi:2024-aqn-glow-pt-1,Sommer:2024-aqn-glow-pt-2}, the AQN emission is dominated by the ionized gas rather than the neutral gas. The reason is that the Coulomb force will enhance the interaction cross-section between AQNs and ionized gas, compared to the AQN-neutral gas cross-section. In the ionized gas scenario, the AQN temperature can be expressed as follows \cite{Majidi:2024-aqn-glow-pt-1}:
\begin{equation}
\label{eq:T_AQN}
T_{\rm AQN}
=\frac{m_ec^2}{k_B}\left[\frac{2{\rm\,GeV\,}(1-g)\hbar}{8\alpha^{3/2}k_B^2}
\frac{3\pi n_{\rm ion}{\rm\Delta v}R^2}{T_{\rm gas,eff}^2}\right]^{\frac{4}{7}}\,,
\end{equation}
where $n_{\rm ion}$ is number density of the ionized gas, $g\approx0.1$ is the fraction of the non-thermal emission, ${\rm \Delta v=|{\bf v}_{DM}-{\bf v}_{bar}|}$ is the relative speed between the AQN and the baryonic matter, and $R$ is the geometrical radius of the AQN. $T_{\rm gas,eff}$ is the effective gas temperature defined as the thermal temperature $T_{\rm gas}$ of the ionized gas plus the kinetic energy (in the gas frame) resulting from the relative motion of dark matter and the baryons:
\begin{equation}
\label{eq:k_B T_gas eff}
k_B T_{\rm gas,eff}
\equiv k_B T_{\rm gas}+\frac{1}{2}m_p{\rm\Delta v}^2\,.
\end{equation}

The effective temperature \( k_B T_{\rm gas,eff} \) determines whether a gas particle will be captured by the Coulomb attraction of the charged AQN. Outside a galaxy, the kinetic term \( \frac{1}{2} m_p \Delta \rm v^2 \) is generally negligible compared to the thermal term \( k_B T_{\rm gas} \), since the relative bulk motion between baryonic and dark matter is small \cite{Majidi:2024-aqn-glow-pt-1, Sommer:2024-aqn-glow-pt-2}. However, this is no longer true in the present study, where we focus on the scenario within the Milky Way's galactic disk. In this case, the baryonic gas rotates with the stellar disk, relative to the non-rotating dark matter halo, where dark matter particles also move randomly. The kinetic term can become larger than the thermal term, especially when the gas is cold. To account for this effect, we introduce the precise definition of \( T_{\rm gas,eff} \) in equation \eqref{eq:k_B T_gas eff}.

% \ls{May suggest to remove footnote 1 and use the new paragraph below}

%\ls{remove last sentence in footnote 5, "When using intensity..." It confuses the referee.}

\edit{Additionally, an alternative emission mechanism was proposed in \cite{Flambaum:2021xub,Flambaum:2021awu,Flambaum:2022wcs}. The authors claim that the spectral emissivity $F_\nu$ has significantly different frequency and temperature dependence. For the relevant energy scale in this work, $k_B T_{\rm AQN}\sim 1{\rm\,eV}$, their computed emission is nearly 100 times larger than that in Eq. \eqref{eqs:F_nu etc.} \cite{Flambaum:2021awu}. This claim is fundamentally incorrect because the total emission intensity is determined by annihilation rate and must be identical for all models of emission. The spectral emissivity may depend on the model, but not the total intensity. The detailed arguments addressing deficiencies of the approach in \cite{Flambaum:2021xub,Flambaum:2021awu,Flambaum:2022wcs} are given in Appendix \ref{app:emission}. Thus, we adopt the original approach of \cite{Forbes:2008uf}  for the spectrum and use Eq. \eqref{eqs:F_nu etc.} in this paper.}

\subsection{From the emission spectrum to the GALEX FUV flux}
\label{subsec:aqn2galexflux}

We expect that the AQN dark matter and ionized gas in a large region surrounding the solar system will interact to form a sky glow, the intensity of which, $I_\nu$  in [$\rm erg\,s^{-1}\,cm^{-2}\,Hz^{-1}\,sr^{-1}$], is defined as:

\begin{equation}
\int_{\Delta \Omega} I_\nu(\hat{\mathbf{n}})\,\rmd\Omega
=\int  \frac{\epsilon_\nu(r,\hat{\mathbf{n}})}{4\pi r^2}\, e^{-\tau_\nu(r,\hat{\mathbf{n}})} \, \rmd r\,r^2\rmd\Omega,
\end{equation}
where $r$ is the distance from the observer to the source in direction $\hat{\mathbf{n}}$, $\epsilon_\nu(r,\hat{\mathbf{n}})$ is the spectral emission per volume in [$\rm erg\,s^{-1}cm^{-3}Hz^{-1}$] and $\tau_\nu(r,\hat{\mathbf{n}})$ is the optical depth. The spectral emission is given by:
\begin{equation}
\label{eq:epsilon_nu}
\epsilon_\nu
=4\pi R^2 n_{\rm AQN} F_\nu\,
\end{equation}
where $F_\nu$ is given by Eq. \eqref{eqs:F_nu etc.}. By taking the derivative with respect to $\Omega$ on both sides, we obtain
\begin{equation}
\label{eq:I_nu hat n}
I_\nu(\hat{\mathbf{n}})
=\frac{1}{4\pi}\int_0^\infty\rmd r\,\epsilon_\nu(r,\hat{\mathbf{n}})e^{-\tau_\nu(r)}\,,
\end{equation}
A full treatment of optical depth is beyond the scope of this work. Instead, we will assume a uniform cutoff at distance $L$. For a wavelength in the range of the GALEX FUV $\lambda\in[1350,1750]$\AA, we adopt the choice of $L=0.6{\rm\,kpc}$, consistent with the estimations done in \cite{Henry:2015-uv-excess-galex}. This assumption is equivalent to modelling optical depth as a step function—allowing full transmission up to $r=L$, with no transmission beyond $r>L$. \edit{While we recognize that a more realistic model for the optical depth would involve an exponential cutoff instead, we maintain the hard cutoff for consistency with \citep{Henry:2015-uv-excess-galex}, as our main goal is to verify the AQN signal using the same tools and approximations.}

In the GALEX analysis \cite{Henry:2015-uv-excess-galex}, the observable is characterized by the photon number flux density $\Phi_\lambda$ [$\text{photon}\, \text{cm}^{-2} \, \text{s}^{-1} \, \text{sr}^{-1} \, \text{\AA}^{-1}$]. The conversion from $I_\nu$ to $\Phi_\lambda$ is given by:
\begin{equation}
\label{eq:Phi_lambda hat n}
\Phi_\lambda(\hat{\mathbf{n}})
=\left.\frac{I_\nu(\hat{\mathbf{n}})}{2\pi\hbar\nu}\frac{\rmd\nu}{\rmd\lambda}
\right|_{\nu=c/\lambda}
=\left.\frac{1}{8\pi^2\hbar\lambda}
\int_0^L\rmd r\,\epsilon_\nu(r,\hat{\mathbf{n}})
\right|_{\nu=c/\lambda}\,.
\end{equation}

The observed averaged flux over all sky is given by:

\begin{equation}
\label{eq:langle Phi_lambda rangle_Delta V}
\langle\Phi_\lambda\rangle
=\frac{1}{4\pi}
\int_0^Lr^2{\rm d}r \oint{\rm d}\Omega\frac{\Phi_\lambda(\hat{\mathbf{n}})}{r^2}
=\frac{1}{32\pi^3\hbar\lambda}\left.
\int_0^Lr^2{\rm d}r\oint{\rm d}\Omega\,\frac{\epsilon_\nu(r,\hat{\mathbf{n}})}{r^2}
\right|_{\nu=c/\lambda}\,.
\end{equation}

Using the relation $\rmd V=r^2\rmd r\rmd\Omega$, Eq. \eqref{eq:langle Phi_lambda rangle_Delta V} can be rewritten in the form of a volume integral:
\begin{equation}
\label{eq:Phi_continuous}
\langle\Phi_\lambda\rangle
=\frac{1}{32\pi^3\hbar\lambda}\left.
\int\rmd V\,\frac{\epsilon_\nu(r,\hat{\mathbf{n}})}{r^2}
\right|_{\nu=c/\lambda}\,.
\end{equation}

In the next section, we will introduce the simulations used in this work; for this reason, it is useful to introduce the discretized version of Eq. \eqref{eq:Phi_continuous},

\begin{equation}
\label{eq:Phi_discretized}
\langle\Phi_\lambda\rangle
=\frac{\Delta r^3}{32\pi^3\hbar\lambda}\left.
\sum_{i,j,k}\,
\frac{\epsilon_\nu(i,j,k)}{r^2}
\right|_{\nu=c/\lambda}\,,  
\end{equation}
where $(i,j,k)$ is the Cartesian coordinate of the voxels in the simulation cube, $\Delta r$ is the voxel width, and $r$ specifies the distance between the $(i,j,k)$-th voxel and the observer (at the Sun location). The sum $\sum_{i,j,k}$ is taken over all voxels with $0<r<L$ and averaged over the full sky.

One final step remains: Eq. \eqref{eq:Phi_discretized} defines the flux $\langle\Phi_\lambda\rangle$ for a specific wavelength $\lambda$. To compare our signal forecast with the measured GALEX FUV excess, we must average over the GALEX FUV bandwidth ($1350$--$1750$ \AA). As a first approximation, we assume a flat filter response function for GALEX. \edit{This simplification is well-motivated, as a flat source spectrum is one of the two benchmark models used by the GALEX collaboration to define the filter's characteristics, such as its effective wavelength~\cite{Morrissey:2007-GALEX-parameters}. Since the AQN emission model is approximately flat in this window, our approach is consistent with a standard characterization of the instrument.} Thus, the AQN signal forecast is given by:
\begin{equation}
\label{eqn:phi_avg_final}
\langle \Phi_\lambda\rangle_{\rm GALEX}
=\frac{1}{\Delta\lambda}
\int_{\lambda_\text{min}}^{\lambda_\text{max}}\dd \lambda \,
\langle\Phi_\lambda\rangle,
\end{equation}
with $\lambda_\text{min}=1350$ \AA, $\lambda_\text{max}=1750$ \AA, and  $\Delta \lambda=\lambda_\text{max} - \lambda_\text{min}$.

%#################################################################|
%#################################################################|
\section{Modelling the GALEX FUV flux}\label{sec:modeling-aqn}

The AQN emission mechanism was presented in section \ref{subsec:aqn-emission}, and section \ref{subsec:aqn2galexflux} outlines the calculation of the GALEX FUV flux. This calculation requires knowledge of the following physical quantities: (i) the DM and baryonic matter densities $n_\text{DM}, n_\text{gas}$, (ii) the relative velocity between DM and baryonic matter, $\Delta\text{v}$, (iii) the gas temperature $T_\text{gas}$, and (iv) the gas ionization fraction $X_\text{e}$. \edit{The AQN spectral emissivity is highly sensitive to variations in gas density, temperature, and relative velocity. These quantities scale nonlinearly with the AQN temperature, as seen in Eq. \eqref{eq:T_AQN}, which in turn scales nonlinearly with the AQN spectral emissivity in Eq. \eqref{eqs:F_nu etc.}. Because of this effect, analytical models which smooth over local variations in these parameters do not have the necessary resolution to estimate the AQN signal accurately. This is why we did not use for instance the Navarro-Frenk-White \citep{Navarro:1996-NFW} or Burkert \citep{Burkert:1995-Burkert} profiles for dark matter and the Disk-Bulge-Halo (DBH) model for visible matter \citep{Widrow:2005-disk-bulge-halo, McMillan:2011-NFW}.}

Since the mean free path of an FUV photon is on the order of $1$--$2$ kpc \cite{Henry:2015-uv-excess-galex} in the disk, we must rely on high-resolution simulations of these physical quantities within a small sub-kpc region centered around the Sun. The advantage of using hydrodynamical simulations is that they allow us to select locations that best represent our Solar System environment and provide a more accurate treatment of the fact that the AQN signal does not scale linearly with these parameters, as demonstrated in section \ref{sec:aqn}. For these reasons, we have chosen to use the FIRE-2 Latte suite \cite{Wetzel:2016-FIRE-Latte}. The following sections detail the process of modelling the FUV signal using these simulations by implementing Eq. \eqref{eqn:phi_avg_final} and the AQN annihilation emission mechanism discussed in section \ref{subsec:aqn-emission}.

\subsection{Milky Way-like galaxy simulations -- FIRE-2 Latte suite}\label{subsec:fire-simulations}

In this work, we utilize the m12i Milky Way-like galaxy simulation, which employs the Feedback in Realistic Environments 2 (FIRE-2) engine. The m12i simulation is part of the FIRE-2 Latte simulation suite\footnote{\href{https://fire.northwestern.edu/latte/}{fire.northwestern.edu/latte/}}, first introduced in \citep{Wetzel:2016-FIRE-Latte}. The Latte project models the formation of Milky Way-mass (MW-mass) galaxies down to redshift 0 within the $\Lambda$CDM cosmological framework, using the FIRE-2 physical model. This model includes detailed treatments of star formation, stellar feedback (such as supernovae, stellar winds, and radiation), and the effects of gas cooling and heating across a broad range of densities and temperatures \citep{Hopkins:2018-FIRE-2-phsyics}.

MW-mass candidate galaxies were selected from a periodic simulation box of 85.5 Mpc in length, based solely on halo mass, to match $M_{200m}=1\text{--}2\cdot 10^{12} M_\odot$. Each chosen MW-mass halo has a particle mass resolution of $7070\,M_\odot$ for gas and $35000\,M_\odot$ for dark matter, with a spatial resolution of 40 pc for dark matter and a minimum of 1 pc for gas (with a median resolution of $25$--$60$ pc at $z=0$). An adaptive softening kernel is employed to achieve higher resolution when necessary. The data for three MW-mass halos from the Latte suite are publicly available\footnote{\href{http://ananke.hub.yt/}{ananke.hub.yt/}}, and the m12i simulation was selected for this project.

\subsection{Pre-processing -- Voronoi tessellation}\label{subsec:voronoi}

In this section, we describe the processing applied to the m12i MW-like galaxy data in preparation for the FUV AQN signal calculation. Since the m12i simulation provides data in the form of particle information, it is necessary to construct a continuous density field. This is achieved by performing a Voronoi tessellation, which converts the discrete particle data into three-dimensional density fields.

As mentioned in section \ref{sec:modeling-aqn}, five physical quantities must be determined locally: $n_\text{DM}$, $n_\text{gas}$, $\Delta \text{v}$, $T_\text{gas}$, and $X_\text{e}$. Along with the AQN mass, these serve as the inputs for estimating the AQN signal, as outlined in section \ref{subsec:aqn-emission}. Before applying the Voronoi tessellation, we align the galactic disk with the xy-plane of our coordinate system to simplify calculations. This alignment is achieved using a weighted linear least squares regression fit to a plane equation applied to the gas particles located at distances of $8\pm1.2$ kpc from the galactic centre (approximately the position of the Sun). From this, the correct rotation vector is determined.

\begin{figure}[ht]
    \centering
    \includegraphics[width=0.99\linewidth]{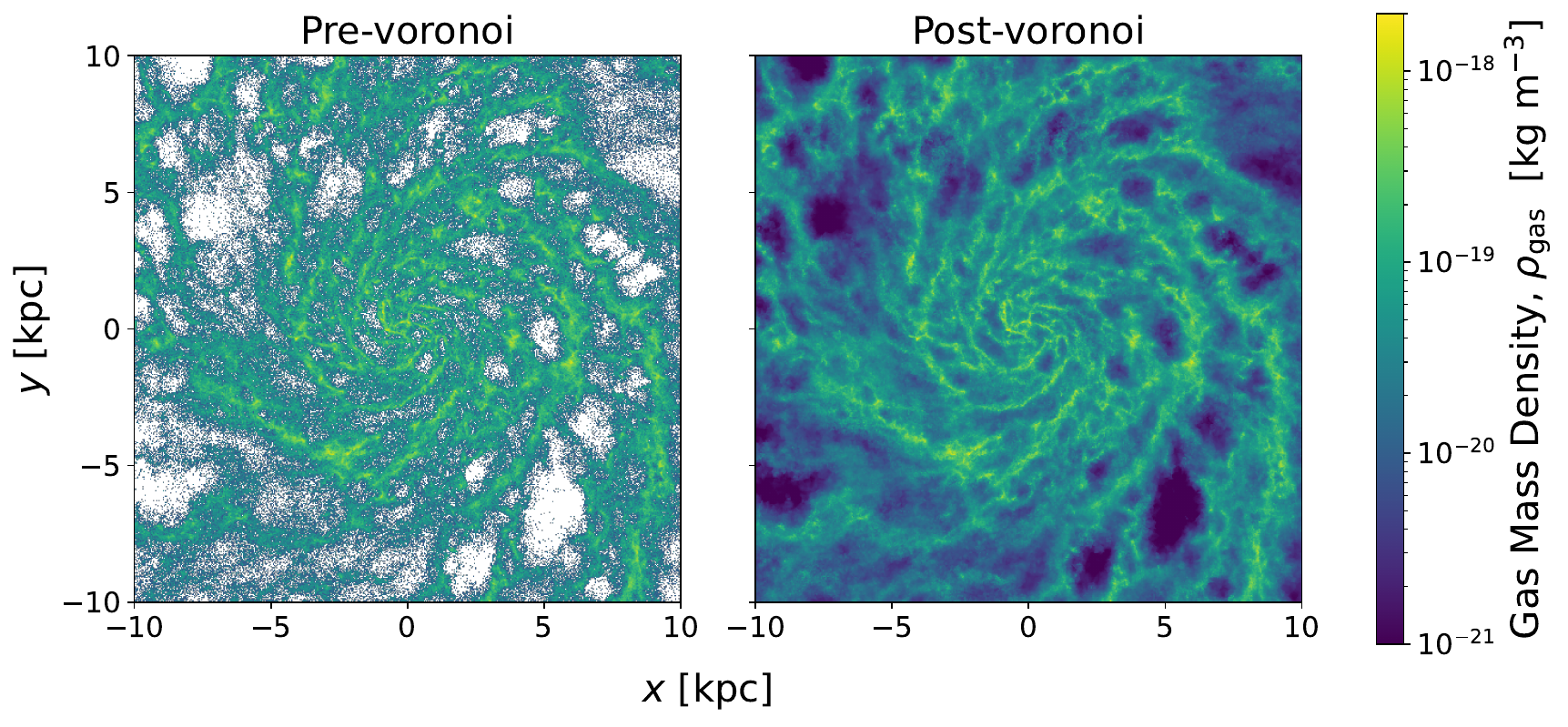}
    \caption{Gas distribution in a 20x20x4 kpc$^3$ cube around the centre of the m12i galaxy simulation. $xy$-plane projections summed over $z\in[-1,1]\;\text{kpc}$. Left: gas mass densities, obtained by binning particles over an overlaid grid. Right: gas mass density field, generated by applying the Voronoi tessellation and nearest-voxel search process discussed in section \ref{subsec:voronoi} to the data in the left plot.}
    \label{fig:voronoi-demo-FIRE}
\end{figure}

A discretized Voronoi tessellation was applied to the rotated m12i particles to convert the particle data into discretized continuous fields. A 20x20x20 kpc cube, centered on the Galactic centre, was selected for this process. The cube was discretized to a resolution of $512$ voxels per side, resulting in each voxel being a cube with a side length of $L_\text{vox}=20 \, \text{kpc} / 512\approx0.039\,\text{kpc}$. Particles were assigned to voxels based on their spatial positions. The choice of this voxel size was driven by a trade-off: At resolutions higher than 512 voxels per side (i.e., cubes with side lengths smaller than 0.039 kpc), the number of empty voxels (those containing no particles) increases significantly in the solar neighbourhood region, which is precisely defined and discussed in section \ref{subsec:solar-neighbourhood}. The benefits of a finer resolution become negligible when compared to the substantial increase in computing time.

For voxels containing no particles, a nearest-neighbour search was performed using a 3D tree search algorithm. This assigned each empty voxel to the nearest non-empty neighbouring voxel, with the particle data from the non-empty voxel applied to all of its empty neighbours, creating a uniform region. The particle data in each region was then transformed into a density field as follows: for mass density, the total mass of all particles in a region was divided by the total region volume. This technique preserved the total mass in the simulated Galaxy. For gas temperature and neutral hydrogen fraction, the weighted average value of all particles in the region was assigned to all voxels within that region. An example of the resulting m12i gas density fields is shown in Figure \ref{fig:voronoi-demo-FIRE}. The mass densities of DM and gas were then converted into number densities using a fixed AQN mass $m_\text{AQN}$ and the assumption that all neutral gas is HI and all ionized gas is HII. This assumption implies that all gas consists of hydrogen, allowing the proton mass to be used in the conversion. With this approach, the number densities $n^{i}_\text{DM},\,n^{i}_\text{gas}$ were obtained for each $i$-th voxel. The total mass of DM and gas components was conserved here, since only the mass of the constituent particles is assumed and modified in this process.

\begin{figure}[ht]
    \centering
    \includegraphics[width=0.6\linewidth]{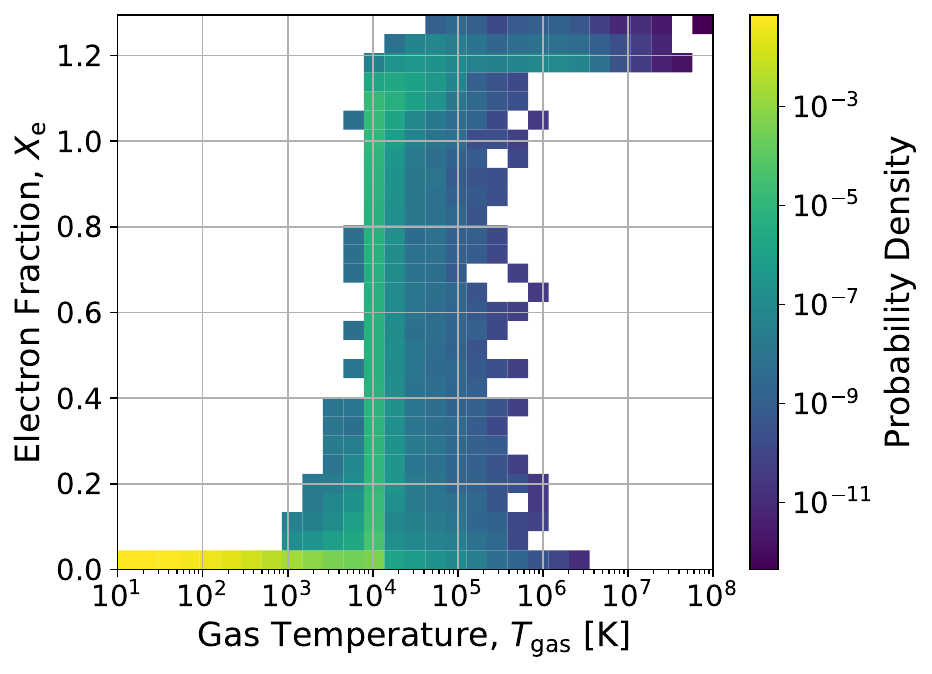}
    \caption{Probability density of the electron fraction $X_e$ and gas temperature $T_{\rm gas}$ for a 20x20x20 kpc cube centered on the Galactic centre of the m12i simulation.}
    \label{fig:efrac_vs_gas_temp}
\end{figure}

As discussed in section \ref{subsec:aqn-emission}, since neutral gas does not contribute significantly to the AQN FUV signal, we only need to estimate the ionized gas number density at each voxel, $n^i_\text{ion}$. Since an average gas temperature is assigned to each gas particle, one possible approach is to use a simple temperature threshold to distinguish between the neutral and ionized gas phases. This method was applied in \cite{Majidi:2024-aqn-glow-pt-1}, where an average ionization temperature of $2.6~\text{eV}\simeq3\times10^4~\text{K}$ was assumed.
However, in our case, this approximation is not valid because multiple physical processes can influence the ionization fraction within a galactic disk. Moreover, given that the AQN signal follows the relation $\Phi\propto T_{\rm gas}^{-26/7}$, it makes the sensitivity to an arbitrary ionization temperature even stronger. Consequently, we opted to use the ionization fraction $X_e$ provided by the m12i simulation. Figure \ref{fig:efrac_vs_gas_temp} presents a 2-D histogram of ionization fraction and temperature for gas particles in the m12i galaxy, illustrating that ionized gas ($X_\text{e} > 0.5$) remains present even at temperatures as low as $\sim 10^4\; K$. Thus, the number density of ionized hydrogen gas in each voxel, $n^{i}_\text{gas}$, was determined based on the ionization fraction $X^i_\text{e}$.

Figure \ref{fig:efrac_vs_gas_temp} shows that $X_e$ exceeds 1, which is expected since the simulation is accounting for ions heavier than hydrogen. For the remainder of this work, we will maintain the simplification that all gas consists solely of hydrogen, allowing $n^i_\text{ion}$ to be calculated as $n^i_\text{ion}=X^i_\text{e} n^i_\text{gas}$. Although this approach artificially increases the total ions number density by $\sim 20\%$, we will show in section \ref{subsec:solar-neighbourhood} that the ionized gas number density from the simulation requires slight scaling to match the conditions of a solar-like region. As a result, at this stage, the uniform $20\%$ overestimate of $n^i_\text{ion}$ is irrelevant.

Each voxel $i$ now contains the following physical quantities: $n^{i}_\text{DM}$, $n^{i}_\text{ion}$, $T^{i}_\text{gas}$. The final required quantity is the relative velocity, $\Delta{\rm v}$ inside each voxel. Although velocity data is available in m12i, we found that its resolution is insufficient, particularly for low-velocity values. The relative velocity between dark matter and baryons is a crucial physical parameter, as evident from Eq. \eqref{eq:T_AQN} and \eqref{eq:k_B T_gas eff}. Specifically, a lower $\Delta{\rm v}$ results in more efficient baryon capture and annihilation by the AQN, therefore it is important to model it carefully. Our approach to incorporating velocity information into our calculations is detailed in section \ref{subsec:velocity-distribution}.

\subsection{Velocity distribution}\label{subsec:velocity-distribution}

\begin{figure}[ht]
    \centering
    \includegraphics[width=0.75\linewidth]{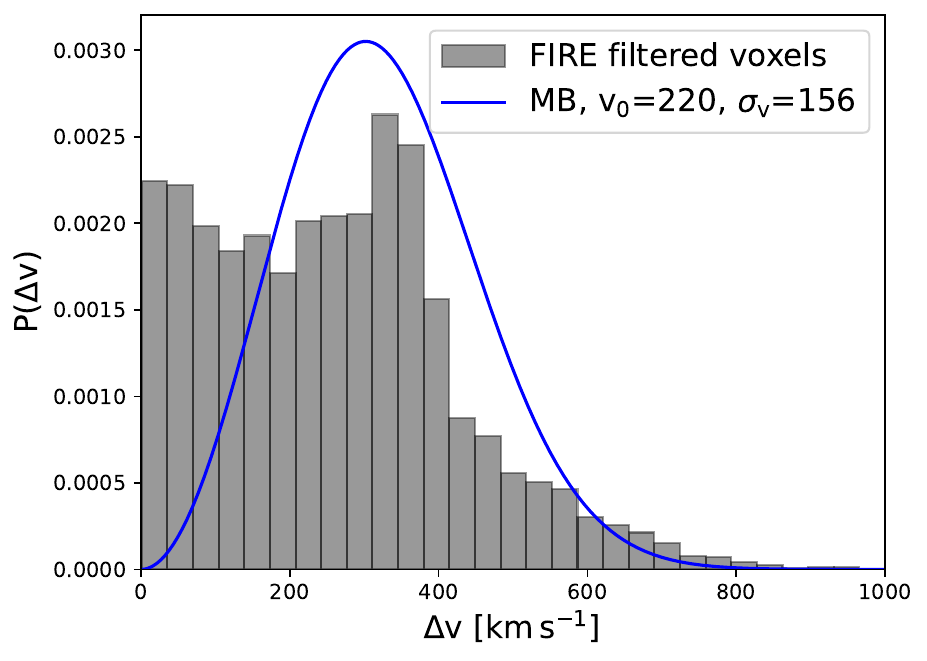}
    \caption{The grey histogram shows the distribution of the relative speed $\Delta \rm{v}$ between dark and visible matter from FIRE's m12i simulation. $\Delta \rm{v}$ values are taken at the solar system neighbourhood locations as described in Section \ref{subsec:solar-neighbourhood}. The solid line shows the shifted Maxwell Boltzmann $\Delta \rm{v}$ model described by Eq. \eqref{eqn:max-boltz-dv}.}
    \label{fig:dv-comparison}
\end{figure}

As described in section \ref{subsec:voronoi}, the physical quantities $n_{\rm AQN}$, $n_{\rm ion}$, $X_e$ and $T_{\rm gas}$ at the voxel location $r,\hat{\mathbf{n}}$ will be used to calculate the spectral emission $\epsilon_\nu(r,\hat{\mathbf{n}})$. These represent the average physical quantities within a voxel. In practice, this should provide a good approximation, except for physical quantities that can vary significantly from one collision to the next, which, in principle, would require resolving individual AQN-baryon collisions.
This is the case for $T_{\rm gas,eff}$ because we approximately have $\epsilon_\nu\propto T_{\rm AQN}^{13/4}\propto T_{\rm gas,eff}^{-26/7}$ from Eqs. \eqref{eqs:F_nu etc.} and \eqref{eq:T_AQN}. It can be seen that for cold gas ($T_{\rm gas} \lesssim \Delta{\rm v}$), it becomes important to treat $\Delta{\rm v}$ properly, particularly for low $\Delta{\rm v}$, as these are the most relevant. This is when we expect the effective AQN-baryon cross-section to be the largest, i.e., the AQN signal is highly sensitive to $\Delta \text{v}$.
Assuming $H(x)$ from Eq. \eqref{eqs:F_nu etc.} is constant, the relationship is approximately $\left<\Phi\right>\sim\Delta \vv^{-5.6}$. The resolution of the per-particle velocity information from m12i is insufficient to describe the tail ends of the velocity difference distribution, which are crucial to the magnitude of the FUV AQN signal. Our solution is to adopt an analytical model for $\Delta \text{v}$ in the Milky Way, using a shifted Maxwell-Boltzmann (MB) distribution. Recent comparisons with both satellite observations (see \citep{Evans:2019-standard-halo-model-dv}) and cosmological simulations (see \citep{Santos:2024-dm-dv}, a follow-up to \citep{Bozorgnia:2016-dm-dv}) show that a shifted MB with a truncated tail models $\Delta \rm{v}$ well. Thus, we choose to model the distribution in relative velocity between DM and VM by:

\begin{equation}\label{eqn:max-boltz-dv}
f(\Delta \text{v}) = 
\sqrt{\frac{2}{\pi}}\frac{\Delta \text{v}^2}{\sigma_{\rm v}^3}
\exp\lr{{\frac{-(\Delta \text{v}^2 + \rm{v}_0^2)}{2\sigma_{\rm{v}}^2}}}
\frac{\sinh\lr{\rm{v}_0 \Delta \text{v} / \sigma_v^2}  }{(\rm{v}_0 \Delta \text{v} / \sigma_v^2)},
\end{equation}
which is a simplified version of the MB distribution from the Standard Halo Model, as summarized in \cite{Evans:2019-standard-halo-model-dv}. Here, ${\rm v}_0=220\;\text{km}/\text{s}$ is the circular rotation speed and $\sigma_{\rm v}={\rm v}_0/\sqrt{2}$, also taken from \cite{Evans:2019-standard-halo-model-dv}. Figure \ref{fig:dv-comparison} shows a comparison between the analytical shifted MB $\Delta \text{v}$ distribution and the $\Delta \text{v}$ distribution obtained by binning FIRE voxels (filtered to solar neighbourhood locations using the process described in section \ref{subsec:solar-neighbourhood}). We note that the binned distribution from the FIRE simulation is skewed towards lower velocities.

% \edit{We note a significant discrepancy between the two distributions in the low-$\Delta v$ regime, which is precisely the region that dominates the AQN flux signal. The distribution from FIRE exhibits an artificial flattening at low velocities, whereas a physically-motivated distribution is expected to have a sharp cutoff proportional to $\Delta v^2$. We argue this is likely a numerical artifact of the Voronoi tessellation method used to process the simulation data, which can have difficulty resolving very small velocity differences. Using the FIRE distribution directly would therefore introduce a significant bias and lead to an underestimate of the AQN signal. We contend that the true physical distribution lies somewhere between the idealized analytical model and the numerically-limited simulation output. For our analysis, we favor the analytical distribution from Eq.~\eqref{eqn:max-boltz-dv}. While it represents an idealized limit, it correctly captures the essential physical behaviour at low velocities, making it a more reliable choice for producing an order-of-magnitude forecast for the FUV excess.}

\edit{We note a significant discrepancy between the two distributions in the low-${\rm \Delta v}$ regime, which is precisely the region that dominates the AQN flux signal. The distribution from FIRE exhibits an artificial flattening at low velocities, whereas a physically-motivated distribution is expected to have a sharp cutoff proportional to ${\rm \Delta v}^2$. We argue this is likely a numerical artifact. Using the FIRE distribution directly would therefore introduce a significant bias and lead to an underestimate of the AQN signal. For our analysis, we favor the analytical distribution from Eq.~\eqref{eqn:max-boltz-dv}. While it represents an idealized limit, it correctly captures the essential physical behaviour at low velocities, making it a more reliable choice for producing an order-of-magnitude forecast for the FUV excess.}

% We claim that the true physical distribution lies somewhere between the idealized analytical model and the numerically-limited simulation output

% In order to avoid the bias that would be introduced by using the FIRE $\Delta{\rm v}$, 

\edit{Thus,} we will use the analytical distribution from Eq. \eqref{eqn:max-boltz-dv}, and the spectral emission we will use in our analysis is given by:

\begin{equation}
\label{eq:langle epsilon_nu rangle}
\overline{\epsilon_\nu(r,\hat{\mathbf{n}})}
= \int\rmd \Delta{\rm v} f(\Delta{\rm v})\,\epsilon_\nu(r,\hat{\mathbf{n}}).
\end{equation}

Thus, we are able to capture the full sensitivity of the FUV AQN signal to variations in $\Delta \text{v}$. Since the AQN signal scales non-linearly to $\Delta \text{v}$, we will see in section \ref{sec:results} that the effect of this on the modelled FUV AQN signal is substantial.

\subsection{Finding solar neighbourhood locations}\label{subsec:solar-neighbourhood}

Given that the optical depth of GALEX FUV photons is of the order of $\sim\,$1 kpc in the solar system region, it is important to select locations in m12i that closely resemble sun-like regions. To make this selection, we will choose regions where the dark matter mass density \( \rho_{\rm DM} \), the ionized and neutral gas fractions, \( n_{\rm ion} \) and \( n_{\rm neut} \), are closest to the measured values at a distance of \( R_\odot \sim 8.2\; \rm kpc \) from the galactic center. Local averages of these parameters from observations are used. The values, tolerance range, and corresponding references are shown in table \ref{tab:solar-sys-params}.

\textbf{\begin{table}[ht]
\centering
\begin{tabular}{@{}lccccc@{}}
\hline
\textbf{Parameter} & \textbf{Value} & \textbf{Range} & \textbf{Unit} & \textbf{Reference} \\ 
\hline
Dark Matter Density ($\rho_{\text{DM}}$) 
& 
0.42 & 0.06 & GeV/cm$^3$ & 
\cite{Staudt:2014-sliding-into-dms}
\\
Ionized Gas Density ($n_{\text{ion}}$) 
& 
0.018 & 0.002 & cm$^{-3}$ & \cite{2008AA...490..179B}
\\
Neutral Gas Density ($n_{\text{neut}}$) 
& 
0.195 & 0.033 & cm$^{-3}$ & \cite{Swaczyna:2020-neut-gas-ref}%\cite{Puyoo_1997}
% \\
% Velocity of Sun ($v_{\odot}$) 
% & 
% 232.8 & 3.0 & km/s & [Ref]%\cite{mcmillan2017}
\\
% Height of Sun above disk ($z_{\odot}$) 
% & 
% $X_{z_{\odot}}$ & $\sigma_{z_{\odot}}$ & pc & [Ref] 
% \\
Radial distance of Sun $(R_{\odot}$) 
& 
8.20 & 0.09 & kpc & \cite{McMillan:2016-radial-dist-ref} %\cite{mcmillan2017}
\\
\hline
\end{tabular}
\caption{Solar region parameters and their respective we used for the selection of solar neighbourhoods regions. The local dark matter density, $\rho_{\rm DM}$, was obtained from the average of twelve recent studies, computed and tabulated in \cite{Staudt:2014-sliding-into-dms}.}
\label{tab:solar-sys-params}
\end{table}}

\begin{figure}[ht]
    \centering
    \includegraphics[width=0.99\linewidth]{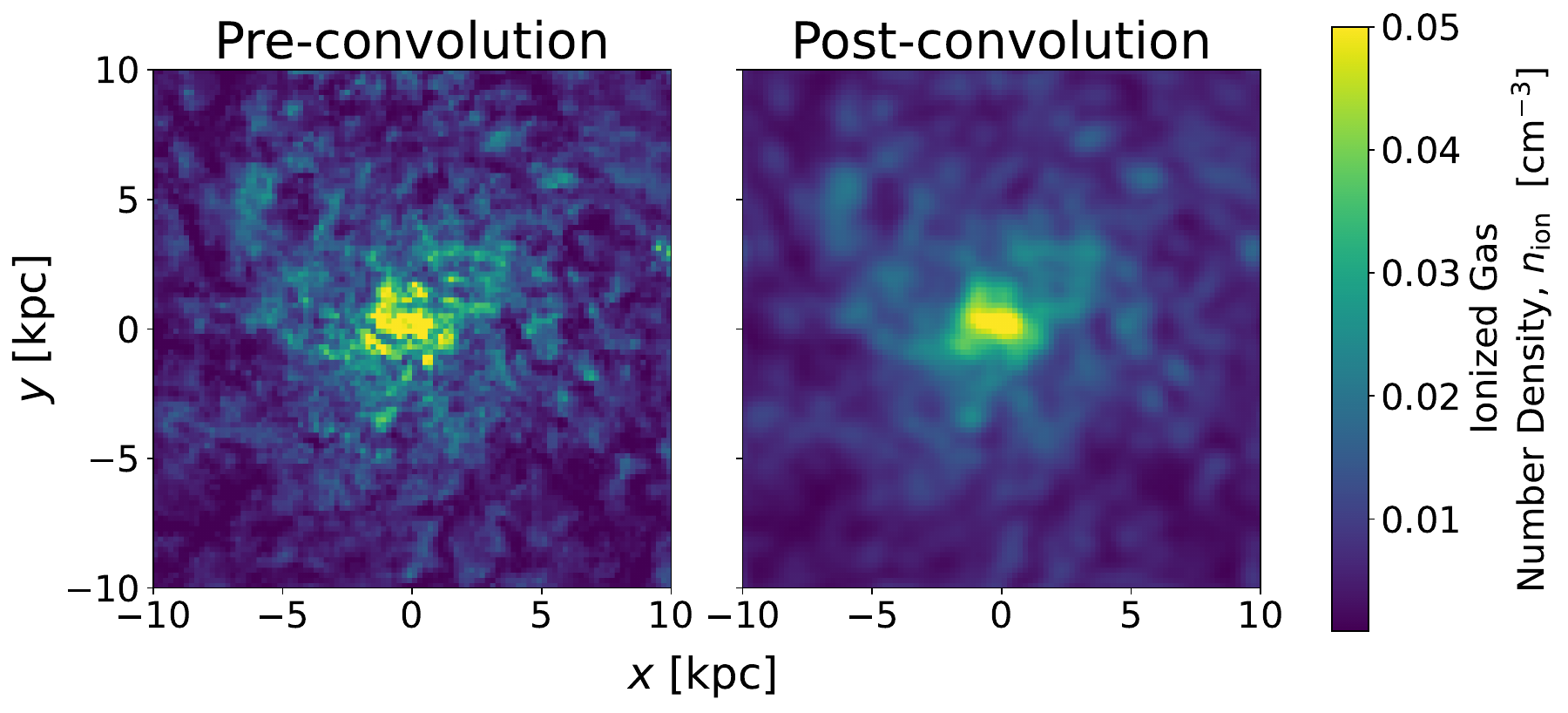}
    \caption{Pre-convolution and post-convolution ionized gas density distributions, using a spherical averaging kernel with $R=0.6\;\text{kpc}$. The $xy$-planes are projections along the $z$-axis, averaged over $z\in[-1,1]\;\text{kpc}$.}
    \label{fig:convolution}
\end{figure}

\begin{figure}[H]
    \centering
    \includegraphics[width=0.6\linewidth]{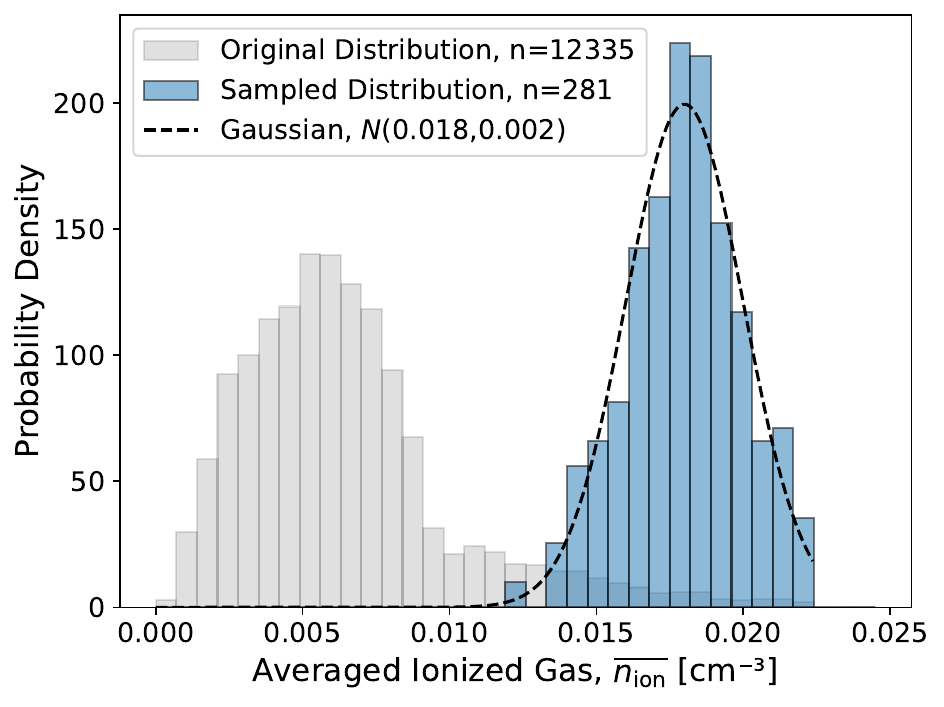}
    \caption{Grey histogram: distribution of the average ionized gas density $\overline{n_{\rm ion}}$ of the central voxel of the solar system regions selected on dark matter density $\rho_{\rm DM}$, neutral gas $n_{\rm neut}$ and distance from the galactic center $R_\odot$ (see Table \ref{tab:solar-sys-params}). The post-convolution of the ionized gas distribution was used to select the central voxel (see right panel of Figure \ref{fig:convolution}). Blue histogram: random sub-sample of the grey histogram such that it follows the gaussian distribution (dashed line) for $n_{\rm ion}$ as indicated by Table \ref{tab:solar-sys-params}. A total of 281 solar system-like regions fullfil all selection criteria of Table \ref{tab:solar-sys-params}.}
    \label{fig:ioni_gas_avg_sampling}
\end{figure}

The sun-like regions were selected as follows: Candidate voxels \( i \) from the 20x20x20 kpc cube were initially selected based on the distance from the galactic centre and the galactic mid-plane criteria, within a shell defined by \( |z_i| < 1\,\text{kpc} \) for the z-coordinates and \( |R_i - R_\odot| < 1.2\,\text{kpc} \) for the radial distances from the centre. We then refined the voxel selection by requiring the dark matter mass density \( \rho_{\rm DM}^i \) and the neutral gas densities \( n_{\rm neut}^i \) to lie within the range specified in table \ref{tab:solar-sys-params}. The use of the ionized gas density \( n_{\rm ion} \) as a selection criterion is slightly complicated by the fact that ionized gas is clumped into many small regions, with an average given by \( n_{\text{ion}} = 0.018 \pm 0.002\,\rm cm^{-3} \) over an average distance of \( 0.93\,\rm kpc \) \cite{2008AA...490..179B,Haffner:2009-ioni-gas-ref}. We therefore decided to apply the \( n_{\rm ion} \) criterion to the FIRE ionized gas distribution, smoothed with a sphere of radius \( 0.6\,\rm kpc \), which corresponds to our approximate mean free path for the GALEX UV photons. This was done to avoid any bias in our selection of solar system candidates that could be caused by local \( n_{\rm ion} \) clumps. The average \( n_{\text{ion}} = 0.018 \pm 0.002\,\rm cm^{-3} \) reported by \cite{2008AA...490..179B} was obtained from 38 pulsars at known distances (ranging from \( 0.2\,\rm kpc \) to $\sim\,$10 kpc), and it is reported that the average ion density is largely independent of the line of sight. We are therefore confident that our averaging procedure is selecting regions representative of the solar system environment. The effect of the \( 0.6\,\rm kpc \) smoothing on the ion gas density is shown in Figure \ref{fig:convolution}, where we see that the pre-convolved gas disk contains many clumps of size \( < 1\,\rm kpc \). The last step is to ensure that our distribution of \( n_{\rm ion} \) \edit{has the same average and standard deviation as given in Table \ref{tab:solar-sys-params} } (\( n_{\text{ion}} = 0.018 \pm 0.002\,\rm cm^{-3} \)), in agreement with \edit{the observations} \cite{2008AA...490..179B}\edit{, which can be accomplished by resampling with a Gaussian:} We noticed that the distribution of \( n_{\rm ion} \) after the selection based on \( R_\odot \), \( n_{\rm neut} \), and \( \rho_{\rm DM} \) has an average \( n_{\rm ion} \) approximately three times smaller than \( 0.018\,\rm cm^{-3} \). This suggests that our solar system might occupy a region that is more ionized than typical regions with similar \( R_\odot \), \( n_{\rm neut} \), and \( \rho_{\rm DM} \). In order to select a sun-like environment, we finalize our selection process by resampling the \edit{post-selection} \( n_{\rm ion} \) \edit{distribution} with a \( {\cal N}(0.018, 0.002) \) Gaussian distribution. The outcome is shown in Figure \ref{fig:ioni_gas_avg_sampling}, where the blue histogram represents the resampled version of the original distribution in gray, where only the \( R_\odot \), \( n_{\rm neut} \), and \( \rho_{\rm DM} \) selection was applied.

% The last step is to ensure that our distribution of \( n_{\rm ion} \) approximates a Gaussian distribution with \( n_{\text{ion}} = 0.018 \pm 0.002\,\rm cm^{-3} \), in agreement with \cite{2008AA...490..179B}. We noticed that the distribution of \( n_{\rm ion} \) after the selection based on \( R_\odot \), \( n_{\rm neut} \), and \( \rho_{\rm DM} \) has an average \( n_{\rm ion} \) approximately three times smaller than \( 0.018\,\rm cm^{-3} \). This suggests that our solar system might occupy a region that is more ionized than typical regions with similar \( R_\odot \), \( n_{\rm neut} \), and \( \rho_{\rm DM} \). In order to select a sun-like environment, we finalize our selection process by resampling the \( n_{\rm ion} \) with a \( {\cal N}(0.018, 0.002) \) Gaussian distribution. The outcome is shown in Figure \ref{fig:ioni_gas_avg_sampling}, where the blue histogram represents the resampled version of the original distribution in gray, where only the \( R_\odot \), \( n_{\rm neut} \), and \( \rho_{\rm DM} \) selection was applied.

\begin{figure}[ht]
    \centering
    \includegraphics[width=0.9\linewidth]{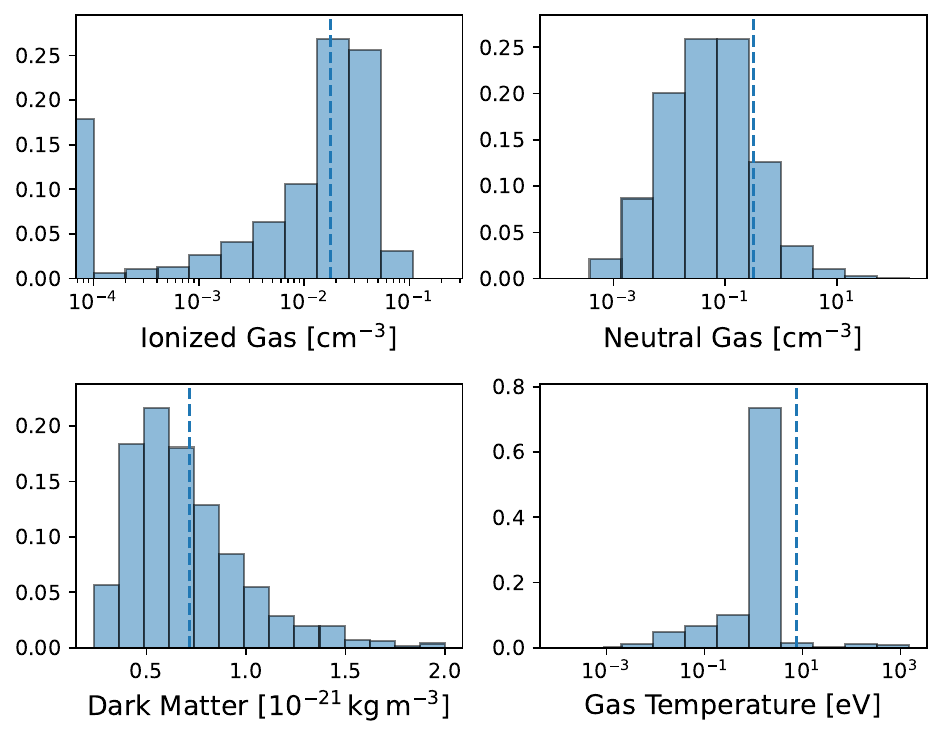}
    \caption{Probability distribution functions for all the voxels of the selected 281 solar system-like regions which are within $0.6\;\rm kpc$ from the central voxel. The dashed blue line shows the mean of the distributions. 
    }
    \label{fig:filtered-parameter-distributions}
\end{figure}

At the end of the selection process, $281$ sun-like regions were selected, each containing approximately 15,203 voxels. Figure \ref{fig:filtered-parameter-distributions} shows the distribution of the physical quantities $n_{\rm ion}^i$, $n_{\rm neut}^i$, $\rho_{\rm DM}^i$ and $T_{\rm gas}^i$ from all the 4,272,043 voxels of these 281 locations. 

\section{Results}\label{sec:results}

The FUV flux generated by AQNs is calculated using the procedure described in section \ref{subsec:solar-neighbourhood}. It is given by Eq. \eqref{eqn:phi_avg_final}, where \( \langle \Phi_\lambda \rangle \) represents the sky-averaged flux at wavelength \( \lambda \), as described by Eq. \eqref{eq:Phi_discretized}, with an integral over both the velocity distribution [Eq. \eqref{eqn:max-boltz-dv}] and GALEX FUV bandwidth, $1350$--$1750$~\AA. This requires the calculation of the spectral emissivity \( \epsilon_\nu \) in each voxel, which in turn requires the computation of the flux \( F_\nu \) and the number density of AQNs, \( n_{\rm AQN} \), as shown in Eq. \eqref{eq:epsilon_nu}. To calculate \( n_{\rm AQN} \) from the dark matter density \( \rho_{\rm DM} \), we assume that AQNs have an average mass \( m_{\rm AQN} \). In reality, it is likely that AQNs have a mass distribution, but this detail is ignored in this work. There is an indication that \( m_{\rm AQN} \) cannot be larger than a few hundred grams, as a larger mass would have left a detectable signature in the small-scale spectrum of the South Pole Telescope (see Figure 12 in \cite{Majidi:2024-aqn-glow-pt-1}). There is also evidence suggesting that macros must be more massive than $\sim\,$$30$--$55$ g, otherwise, they would have left detectable lattice defects in ancient mica \cite{DeRujula:1984axn,2015MNRAS.450.3418J}. However, the mica constraint is cross-section dependent and assumes all macros have the same mass \cite{2015MNRAS.450.3418J}. This lower bound does not apply to AQNs because AQNs have a mass distribution, and it is uncertain how AQNs would interact chemically with mica. \edit{The 5 g lower bound on the AQN mass is established instead from observations by the IceCube Neutrino Observatory, as described in \cite{Zhitnitsky:2021-aqn-recent-review}}. For our calculations, we choose an AQN mass range of \( m_{\rm AQN} = [10, 100, 500]\, \rm g \).

% mica applies to single-mass DM particles

% here we have mass distribution, which means mica bound is lower than the 55 g.

% can site [17] instead of 1905.00022, there's a limit there.

\begin{figure}[h]
    \centering
    \includegraphics[width=0.998\linewidth]{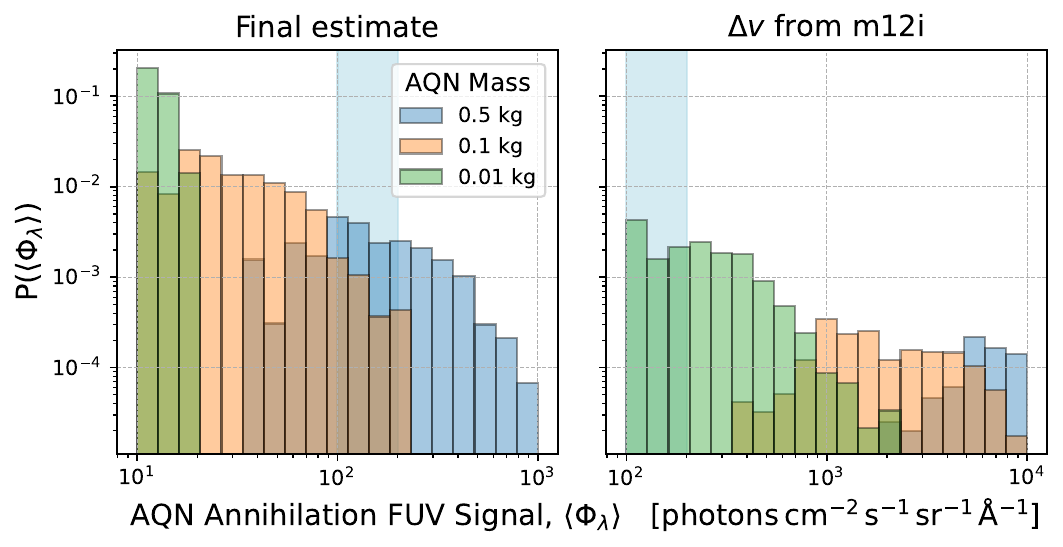}
    \caption{\editt{Probability distributions $P(\langle\Phi_\lambda\rangle)$ of AQN FUV signal from voxels surrounding the candidate solar neighbourhood regions for three AQN masses, with $\Delta \vv$ computed in two different ways: (left) our final estimate, which models $\Delta \vv$ with a Maxwell Boltzmann distribution, and (right) an estimate with $\Delta \vv$ taken from voxels in FIRE's m12i simulation. The GALEX FUV excess (100-200 photon units) is shown as a light blue band. 20 logarithmically-spaced bins are used for each mass.}}
    \label{fig:aqn-signal-histogram}
\end{figure}

\editt{As an intermediate result, we first plot the AQN annihilation FUV signal in the form of a probability distribution: Figure \ref{fig:aqn-signal-histogram} shows histograms of $\langle\Phi_\lambda\rangle$ computed from the voxels surrounding the candidate solar neighbourhood regions in Section \ref{subsec:solar-neighbourhood}. To clearly illustrate the probability of the AQN FUV signal matching or exceeding the GALEX FUV excess, we choose to represent $\langle\Phi_\lambda\rangle$ as a complementary cumulative distribution (CCDF) for our final result. }

\begin{figure}[h]
    \centering
    \includegraphics[width=0.99\linewidth]{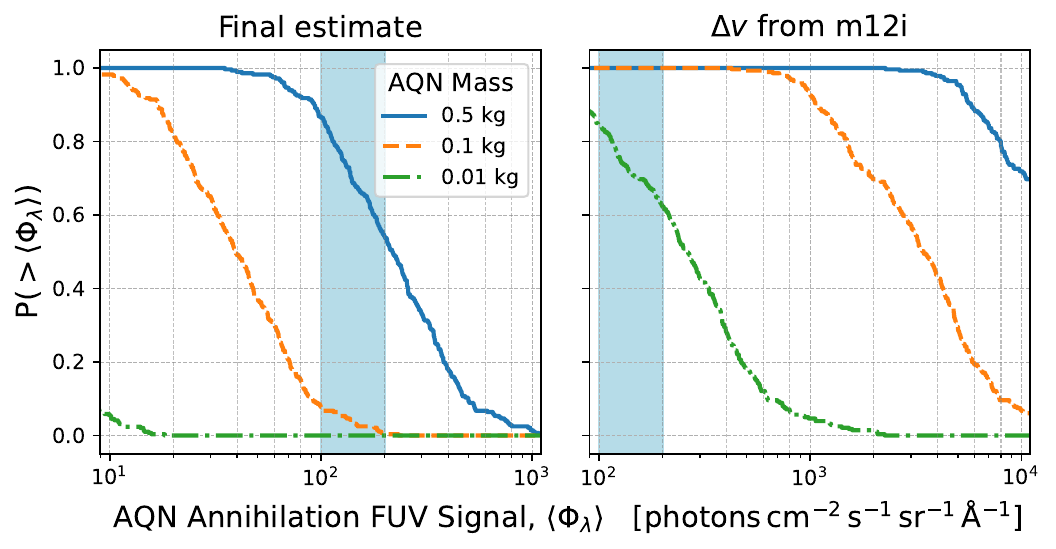}
    \caption{Complementary cumulative distribution function (CCDF) $P(>\langle\Phi_\lambda\rangle)$ of AQN FUV signal from the voxels surrounding the candidate solar neighbourhood regions (selected in Section \ref{subsec:solar-neighbourhood}) for three AQN masses (i.e probability that the signal exceeds a given $\left<\Phi_\lambda\right>$), in two scenarios: (left) our final estimate, using the Maxwell Boltzmann distribution for $\Delta \vv$ (\editt{S}ection \ref{subsec:velocity-distribution}), and (right) an estimate using the $\Delta \vv$ data from voxels from FIRE's m12i simulation. The magnitude of the GALEX FUV excess (100--200 photon units) is shown as a light blue band.}
    \label{fig:aqn-signal-cumulative}
\end{figure}

The left panel on Figure \ref{fig:aqn-signal-cumulative} shows the \editt{CCDF} $P(>\langle\Phi_\lambda\rangle)$ as a function of $m_{\rm AQN}$. The vertical blue band shows the order of magnitude of the measured GALEX FUV excess \cite{Henry:2015-uv-excess-galex,Akshaya:2019-uv-excess}, recently confirmed by the New Horizon spectrograph Alice \cite{Murthy:2025-uv-excess-new-horizons}. The AQN prediction intersects the FUV excess for \editt{an} AQN mass in the expected range ([$100$--$500$] g), if AQNs exist. \editt{This range estimate is obtained using the spectrum in Eq. \eqref{eqs:F_nu etc.} from \cite{Forbes:2008uf}}\footnote{\editt{In Appendix \ref{app:compare-models}, we make a comparison between this and the incorrect heavier mass range estimate obtained if using the spectrum in \cite{Flambaum:2021xub,Flambaum:2021awu,Flambaum:2022wcs}, whose deficiencies are discussed in detail in Appendix \ref{app:emission}}.
}.

We should emphasize that our prediction contains no free parameters, other than the AQN mass. All other physical parameters ($n_{\rm ion}$, $\rho_{\rm DM}$, $T_{\rm gas}$) are obtained from the FIRE simulation with the requirement that they match our solar system environment within $\sim$$1\;\rm kpc$, and the velocity difference $\Delta{\rm v}$ from the expected microphysical velocity distribution.

This is a very encouraging result; however, we should keep in mind that the AQN temperature \( T_{\rm AQN} \) [Eq. \eqref{eq:T_AQN}], which is the main quantity controlling the AQN emission spectrum, has a strong non-linear dependence on some physical parameters. Using physical parameters from a realistic simulation reduces this dependence, but, for instance, the relative velocity \( \Delta \rm v \) is poorly described by the simulation, and we had to rely on Eq. \eqref{eq:T_AQN}. The right panel of Figure \ref{fig:aqn-signal-cumulative} shows what happens to the CCDF when using the incorrect \( \Delta \rm v \) from FIRE (see Section \ref{subsec:velocity-distribution}). The signal increases significantly, exceeding the measured excess by at least one order of magnitude. Nevertheless, we believe that our estimate is the correct order of magnitude and confirms that the AQN model is capable of producing an FUV signal in agreement with the measured excess. 
To our knowledge, this is the only dark matter model capable of producing such an FUV signal. Other models such as WIMPs, QCD axions and primordial black holes are off by at least five orders of magnitude \cite{Henry:2015-uv-excess-galex,Akshaya:2018-uv-excess}.

Future extensions of this work will focus on two aspects. One aspect is to calculate the precise spectrum of the signal. Recent UV observations by New Horizons with the Alice spectrograph \cite{Murthy:2025-uv-excess-new-horizons} have measured an excess of flux in the range 912-1100 \AA, consistent with GALEX FUV \cite{Henry:2015-uv-excess-galex}. In our work, we estimated the AQN flux using a hard cutoff of 0.6 kpc to account for absorption, assuming a constant optical depth near the FUV GALEX bandwidth range,  1350-1750 \AA. As $\lambda$ approaches the ionization limit of 912 \AA, the extinction becomes stronger, and our assumption of constant extinction no longer holds. Below 912 \AA, the photoionization cross-section is increased by a factor $\sim 10^7$ \cite{1996ApJ...465..487V}, leading to a complete suppression in the observed flux. Estimating the wavelength dependent optical depth $\tau_\lambda$ in this part of the spectrum is challenging due to the strong $\lambda$ dependence of the absorption and the inhomogeneous distribution of dust and gas clouds. Fortunately, significant recent progress has been made in this area, which leads to the second aspect for future work. Using the 3-dimensional information of the dust and H$_\alpha$ enabled by the Gaia observations \cite{2024A&A...685A..82E,2025MNRAS.540L..21M}, it should indeed be possible to calculate the FUV emission from AQNs for specific lines-of-sight. Therefore, it should in principle be possible to compare such calculations directly to the signal measured from the 25 fields observed with the Alice spectrograph \cite{Murthy:2025-uv-excess-new-horizons}.

\section{Conclusion and Discussion}\label{sec:discussion}

In this paper, we investigated the impact of the AQN dark matter model on the FUV background in the Milky Way. We found that the annihilation of baryons with antimatter AQNs generates a diffuse FUV background with a photon mean free path on the order of \( \sim 1 \, \text{kpc} \). This background is consistent with the FUV excess measured by GALEX \cite{Henry:2015-uv-excess-galex} and the Alice spectrograph aboard New Horizons \cite{Murthy:2025-uv-excess-new-horizons}. To work with realistic quantities in the region surrounding the Sun, we used the FIRE hydrodynamical simulations. The physical environment has no free parameters (all parameters are set by the simulation), and the AQN model has only one free parameter: the AQN mass. We found that the FUV excess can be reproduced with an average AQN mass that is consistent with the non-detection of the AQN signal by the South Pole Telescope \cite{Majidi:2024-aqn-glow-pt-1}. We argue that the puzzling features discussed in Section \ref{sec:introduction} can be naturally explained within the proposed AQN-induced mechanism.

Our work lays the foundation for a potential new source of FUV photons that is not associated with O-type stars or other known sources, such as B-type stars, compact objects, accretion disks, or cosmic rays. It was suggested in \cite{Henry:2018yar} that a new process, emitting "a continuum of photons in the range of approximately 850 \AA ~ to about 2000 \AA," may be necessary to explain the FUV measurements obtained by the Voyager spectrometers. Similarly, it has been known for some time that "beyond the resolved galaxies, there is an Extragalactic Background Light (EBL) component that cannot be explained by diffuse galaxy halos or intergalactic stars," with a discrepancy that could be as large as a factor of approximately $2$--$3$ \cite{Mattila:2019ybk}. The AQN model, and its potential contribution to the EBL, may help address this gap, as its spectrum is very broad and includes optical and IR light.

The possibility of another source of FUV and ionizing photons could have important consequences for cosmology, in particular on the reionization of the Universe and the formation of the first galaxies. Recent observations from the James Webb Space Telescope (JWST) reveal that early, faint galaxies are prolific producers of ionizing photons. If UV-producing stars are assumed to be the source of these photons, this would imply an early and accelerated reionization of the universe, which would make the optical depth inconsistent with the measurements from Planck \cite{Munoz:2024-JWST-crisis}.
A key element in this discussion is the escape fraction $f_\text{esc}$, the fraction of ionizing photons that escape from galaxies to contribute to reionization, and the ionizing efficient of each galaxy $\xi_{\rm ion}$. According to \cite{Munoz:2024-JWST-crisis}, JWST observations shows that reionization could start as early as $z\simeq 12$ and end at $z\simeq 8$, leading to an optical depth $\tau\sim 0.08$ inconsistent with $\tau_{\rm CMB}=0.055$. The AQN model presented here can produce UV photons consistent with JWST observations, but without the need for an early star formation. The near-Bremsstrahlung emission spectrum given by Eq. (\ref{eqs:F_nu etc.}) can indeed produce UV photons without any higher energy radiation. Moreover, since the AQN signal is diffuse, coming from the interaction between baryons and anti-matter AQNs, it is happening everywhere in the halo, and does not have to be confined in the disk in star formation regions. For this reason, the AQN model can alter the canonical view on the reionization of the Universe and independently impact $f_\text{esc}$ and $\xi_{\rm ion}$. Whether this can be sufficient to explain the JWST result without a significant change in optical depth remains to be demonstrated.

\acknowledgments

We would like to thank Jayant Murthy, Marc-Antoine Miville-Desch\^ene and Arif Babul for discussions at different stages of this work.
MS, XL, FM, BS, LVW, and AZ acknowledge the support from NSERC. All calculations in this work were carried out using the Canadian Advanced Network for Astronomical Research (CANFAR) Science Portal\footnote{\href{https://www.canfar.net/en/}{canfar.net/en/}.} computing system, and we would like address a special thanks to Sebastien Fabbro for his continuous support on CANFAR. Computations were carried out using open-source python packages 
\texttt{numpy} \cite{2020Natur.585..357H}, 
\texttt{scipy} \cite{2020NatMe..17..261V}, 
\texttt{astropy} \cite{2018AJ....156..123A}, and 
\texttt{matplotlib} \cite{2007CSE.....9...90H}.

\appendix

\section{Comments on an alternative emission spectrum }
\label{app:emission}

Recently, an alternative emission spectrum for the same AQN model used in this paper was proposed by Flambaum and Samsonov (FS) \cite{Flambaum:2021xub,Flambaum:2021awu,Flambaum:2022wcs}. The authors claim \cite{Flambaum:2021awu} that the radiation produced by AQNs is 100 times stronger than the original calculation \cite{Forbes:2008uf} by Forbes and Zhitnitsky (FZ), for a typical AQN temperature of $T_{\rm AQN}\sim 1\;\rm eV$. Moreover, FS also argue that the spectrum should exhibit a steep decrease at low frequencies, while the FZ spectrum remains approximately flat, as expected because of the optically thin nature of the Bremsstrahlung emission process.

In this appendix, we provide several reasons why the FS claims are incorrect, explaining why their estimate should not be used and why we should instead adhere to the original derivation. We should emphasize, before discussing the physical details of the emission process, that the factor 100 discrepancy is based on a fundamentally incorrect calculation. The total energy (flux) released from the annihilation should be exactly the same in the FS and FZ model, because it is only determined by the collision rate and does not depend by which mechanism the energy is released \footnote{The discrepancy in intensity between the formulas in \cite{Flambaum:2021xub,Flambaum:2021awu,Flambaum:2022wcs} arises because the temperature $T_{\rm AQN}$ in their approach is an effective parameter based on observable properties like environmental density, which they use as a black body temperature. In contrast, the equivalent $T_{\rm AQN}$ in the FZ approach differs by a factor of 3, leading to a factor of $3^4\simeq 10^2$ discrepancy claimed by FS. When using intensity as a function of observable parameters, such as environmental density, both approaches yield identical results.}.
 
The main difference between the two proposed mechanisms lies not in the total intensity but in the spectral features. In the following subsections, we critically review the proposed FS mechanism to understand the nature of these differences. We conclude that the original spectrum \cite{Forbes:2009wg} provides a more accurate description of the complex physics behind the emission from the AQN's surface.

\subsection{Overview of FS's thermal radiation model}
In \cite{Flambaum:2021awu}, FS considered an antimatter AQN can be described by the Drude model \cite{ashcroft:1993solid}. The Drude model describes the conduction electrons in the metal as the classical ideal gas filling uniformly inside a uniform background of ions. 
The dielectric function is given by \footnote{\label{footnote:convention}We follow the convention of FS in \cite{Flambaum:2021awu} and use natural units $\hbar=c=\epsilon_0=1$.}:

\begin{equation}
	\label{eq:epsilon omega}
	\epsilon(\omega)
	=1-\frac{\omega_p^2}{\omega^2+i\omega/\tau}\,,
\end{equation}
where $\omega_p\approx2{\rm\,MeV}$ is the plasma frequency and $\tau=(475{\rm\,eV})^{-1}$ is the Drude damping time.\footnote{In the original work \cite{Flambaum:2021awu} by FS, the notation is the inverse damping time $\gamma\equiv\tau^{-1}$, namely the collision rate. We choose $\tau$ as the parameter here and reserve $\gamma$ for the relativistic Lorentz factor in the following subsection.} 
FS focused on the dense ultra-relativistic positrons in the electrosphere of an antimatter AQN with a chemical potential $\mu=33.5{\rm\,MeV}\gg m_e$ \cite{Flambaum:2021xub,Flambaum:2021awu}. 
Both parameters, $\omega_p$ and $\tau$, were analytically estimated by FS in the ultra-relativistic limit. Specifically, $\omega_p=\sqrt{\frac{4\pi\alpha}{3}\frac{\partial n_e}{\partial\mu}}$ was obtained from \cite{Delsante:1980dro},  where $\alpha$ is the fine structure constant, $n_e$ is the positron number density, and $m_e$ is the electron mass. $\tau$ was evaluated assuming the positrons scattered off particular antiquarks within the entire quark nugget. The scattering was described by a Thomas-Fermi screening potential \cite{ashcroft:1993solid}, with an inverse screening length $\lambda_D^{-1}\sim3.2{\rm\,MeV}$.

The thermal radiation from an AQN can be produced by oscillations of the electrosphere, characterized by the dielectric function \eqref{eq:epsilon omega}. The emission is conventionally expressed in reference to blackbody radiation $F_{\rm BB}(\omega,T)$ (see e.g. \cite{bohren:1998absorption}):
\begin{equation}
	\label{eq:dF FS omega T}
	\frac{\rmd}{\rmd\omega}F^{\rm (FS)}(\omega,T)
	=E(\omega)\frac{\rmd}{\rmd\omega}F_{\rm BB}(\omega,T)\,,\qquad
	\frac{\rmd}{\rmd\omega}F_{\rm BB}(\omega,T)
	=\frac{1}{4\pi^2}\frac{\omega^3}{\exp(\omega/T)-1}\,,
\end{equation}
where $E(\omega)$ is the emissivity coefficient, which depends on the radius $R$ of the AQN. The explicit form of $E(\omega)$ is defined by an infinite sum of Riccati-Bessel functions parameterized by $\omega R$ (see \cite{Flambaum:2021awu} and the textbook derivation \cite{bohren:1998absorption}). In the long wavelength limit ($\omega R\ll1$), $E(\omega)$ can be appropriated in the following simple form:

\begin{equation}
	\label{eq:E omega approx}
	E(\omega)
	\approx6[1+(\omega R)^2]\,{\rm Re}[\frac{1}{\sqrt{\epsilon(\omega)}}]
	\approx\frac{3}{\omega_p}\sqrt{\frac{2\omega}{\tau}}\,\qquad
	(\omega R\ll1\,,~|\sqrt{\epsilon(\omega)}|\gg1)\,,
\end{equation}
where the refractive index $|\sqrt{\epsilon}|$ is assumed to be very large if $\omega<\omega_p/10$. As argued in \cite{Flambaum:2021awu}, this simple form in fact holds in almost the entire range ($\omega\ll\omega_p$) with a small modification of the numerical prefactor. 

The Drude model is a well-known classical and phenomenological model of metals, designed to describe an ideal gas of non-relativistic electrons. FS applied this model to the ultra-relativistic region of the AQN electrosphere, using an explicit theoretical calculation of the unphysical model parameters (such as positron density and the "damping constant" in their paper \cite{Flambaum:2021awu}), which is incorrect. The necessary quantum and relativistic corrections were never addressed in \cite{Flambaum:2021xub,Flambaum:2021awu,Flambaum:2022wcs}, nor was the expected strong variation of the chemical potential and positron density in the electrosphere with distance from the core considered, nor the presence of an electrosphere boundary. We discuss these corrections in subsections \ref{subsec:Quantum and relativistic corrections} and \ref{other corrections}. These are unavoidable physical phenomena underlying the AQN emission process and must be accounted for in any proper treatment of the problem.

\subsection{Quantum and relativistic effects}
\label{subsec:Quantum and relativistic corrections}

The Drude model is characterized by two phenomenological parameters: the relaxation time $\tau$ and the density $n_e$ of conduction electrons, which determines $\omega_p$. In standard applications of the Drude model to condensed matter physics (e.g., \cite{solyom:2008fundamentals,ashcroft:1993solid}), these parameters are determined experimentally; they are not fundamental parameters and cannot be computed from first principles. However, FS estimated them using erroneous assumptions, which we believe is inappropriate for the AQN model, where unknown physics is involved and no experiments can be conducted to estimate these parameters. In contrast, the approach used in FZ is based on Quantum Electrodynamics and the mean field approximation, both of which are grounded in strong theoretical foundations, and no phenomenological parameters are introduced. In the remainder of this subsection, we provide an estimate of the quantum corrections to the Drude model as applied to AQNs and discuss its limitations.

The quantum generalization of the Drude model is the Lindhard theory (also known as the random phase approximation, see e.g. \cite{solyom:2008fundamentals,ashcroft:1993solid}). The Lindhard dielectric function is evaluated from the first order perturbation of the Schr\"odinger equation:

\begin{equation}
	\label{eq:epsilon q omega}
	\begin{aligned}
		\epsilon(\omega)
		&=1-\frac{4\pi\alpha}{q^2}\Pi(\mathbf{q},\omega)\,,\\
		\Pi(\mathbf{q},\omega)
		&\equiv\frac{\delta n_e(\mathbf{q},\omega)}{U(\mathbf{q},\omega)}
		=\int\frac{\rmd^3k}{4\pi^3}
		\frac{f_0(\varepsilon_{\mathbf{k}})-f_0(\varepsilon_{\mathbf{k+q}})}
		{\omega-\varepsilon_{\mathbf{k}}+\varepsilon_{\mathbf{k+q}}}\,,
	\end{aligned}
\end{equation}
where $\Pi(\mathbf{q},\omega)$ is the proportionality factor (also called the ``response function'') between the charge induced $\delta n_e$ in an interacting electron (or positron) system and the external perturbing potential energy $U(\mathbf{q},\omega)$, the interaction of the electrons (or positrons) is described by the scattering momentum $\mathbf{q}$ and energy $\omega$, $\varepsilon_{\mathbf{k}}=\sqrt{m_e^2+k^2}$ is the free electron (or positron) energy, $f_0(\varepsilon)=1/[\exp(\frac{\varepsilon-\mu}{T})+1]$ is the equilibrium Fermi distribution. 

In the ultra-relativistic limit of the positrons and assuming $q\ll p_F\sim k$ (i.e. positrons only scatter near the Fermi surface), the Lindhard dielectric function becomes:
\begin{equation}
	\label{eq:epsilon q omega approx}
	\epsilon(q,\omega)
	\approx1+\frac{3\omega_p^2}{q^2}\left[
	1-\frac{\omega}{2q}\tanh^{-1}\left(\frac{2q\omega}{q^2+\omega^2}\right)
	\right]\,.
\end{equation}
In the case of metals, the conduction electrons (or positrons in our case)  are approximately free, so the scattering momentum is small, i.e. $\omega\gg q$. In this limit, the Lindhard dielectric function \eqref{eq:epsilon q omega approx} reduces to the Drude one \eqref{eq:epsilon omega} if the relaxation time $\tau$ is small.\footnote{When $\tau$ is also taken into account, one finds the so-called ``Lindhard-Mermin form'' for the dielectric function (see e.g. \cite{solyom:2010fundamentals}):
	\begin{equation*}
		\epsilon(\mathbf{q},\omega)
		=1-\frac{4\pi\alpha}{q^2}
		\frac{[1+i/(\omega\tau)\Pi(\mathbf{q},\omega+i/\tau)]}{
			1+i/(\omega\tau)[\Pi(\mathbf{q},\omega+i/\tau)/\Pi(\mathbf{q},0)]
		}\,.
	\end{equation*}
	It reduces to the form \eqref{eq:epsilon omega} of the Drude model in the limit of $\omega\gg q$, as expected.
}
In the low frequency limit ($\omega\ll q$), the conduction electrons (or positrons) are approximately static. The Lindhard dielectric function reproduces the Thomas-Fermi screening. 

The first limitation of the Drude model is as follows: from a quantum mechanical perspective, the Drude model is only valid in the large \( \omega \) limit. It fails in the small \( \omega \) limit, when the screening effect becomes dominant. For conduction electrons (or positrons) scattering near the Fermi surface, a reasonable range for the scattering momentum is \( q \lesssim 0.1 p_{\rm F} \). In FS's analysis, the Fermi momentum is large, i.e., \( p_{\rm F} \sim \mu = 33.5 \, \text{MeV} \), so \( q \lesssim \text{a few} \, \text{MeV} \). We conclude that FS's estimation of the AQN thermal radiation may remain valid only for sufficiently large \( \omega \gtrsim 1 \, \text{MeV} \), where the screening effect is negligible \footnote{Note that this argument also holds when we apply the Drude model to conventional metals. In the non-relativistic limit, Lindhard dielectric function in case of $\omega\gg q$ becomes:
	\begin{equation*}
		\epsilon
		\approx1-\frac{\omega_p^2}{\omega^2}
		\left[1+{\cal O}(\frac{\varepsilon_{\rm F}q^2}{m_e\omega^2})\right]\,,
	\end{equation*}
	where $\omega_p^2=\frac{4\pi\alpha n_e}{m_e}$ in the non-relativistic limit, and $\varepsilon_{\rm F}\sim\,$eV is the Fermi energy of metals. If we estimate $\frac{q^2}{m_e}\lesssim0.1\varepsilon_{\rm F}\sim0.1\,$eV for metals, we find the Drude model remains valid when $\omega\gtrsim0.1\,$eV. This agrees with the observation, see e.g. \cite{bohren:1998absorption}, the textbook cited by FS in Ref. \cite{Flambaum:2021awu}.
}. 
In comparison, the original treatment by FZ, based on microscopic fundamental physics, remains valid even in the low-frequency limit. 

Even in the large \( \omega \) limit (\( \omega \gg q \)), FS's treatment still requires justification, which was not addressed in \cite{Flambaum:2021xub,Flambaum:2021awu,Flambaum:2022wcs}. According to \cite{Delsante:1980dro}, the FS estimate of the effective plasma frequency \( \omega_p \) turns out to be significantly larger when \( \omega \gg q \gtrsim \frac{1}{14} p_{\rm F} \). This modification, related to the effects of pair production and transverse interactions \cite{Delsante:1980dro, Jancovici:1962otr}, dramatically complicates the calculations. Qualitatively, we expect \( \omega_{\rm p,eff} \sim \gamma \omega_p \). Since \( E(\omega) \propto \omega_p^{-1} \) based on FS's estimation \eqref{eq:E omega approx}, the emissivity coefficient is suppressed by a large Lorentz factor, \( \gamma \sim \frac{p_{\rm F}}{m_e} \sim 68 \). Furthermore, the relaxation time \( \tau \) is estimated based on the assumption of a Thomas-Fermi screening potential. It is inappropriate to use one limit (\( \omega \gg q \)) in the model while applying the opposite limit (\( \omega \ll q \)) to estimate the model's parameters. In summary, any theoretical estimation of phenomenological parameters is not suitable for use in the Drude model as implemented in \cite{Flambaum:2021xub,Flambaum:2021awu,Flambaum:2022wcs}.

\subsection{Other potential corrections: finite-size effect, metallic vs. atomic model}\label{other corrections}

In addition to the quantum and relativistic corrections not addressed by \cite{Flambaum:2021xub,Flambaum:2021awu,Flambaum:2022wcs}, there are other potential problems with FS's treatment, as shown below.

At a very low frequency $\omega$, the emission proposed by FS will be suppressed by a factor of $(\omega R)$ due to the finite-size effect. This occurs when the photon wavelength $\lambda\propto\omega^{-1}$ is longer than the size of the nugget $R$, the suppression factor $(\omega R)\ll1$ becomes effective. For a typical AQN radius of size $10^{-5}{\rm\,cm}$, this suppression effect happens at $\omega\lesssim0.1{\rm eV}$. The ``finite size effect'' refers to the photon wavelength that is so long that we have to deal with the boundary effect (i.e. the shape of the nugget). In the low frequency limit ($\omega R\ll1$), the refractive index $|\sqrt{\epsilon(\omega)}|$ is smaller than assumed by FS [see Eq. \eqref{eq:E omega approx}] due to the presence of the screening effect \eqref{eq:epsilon q omega approx}. There is a soft cutoff at $\omega\lesssim1\,$MeV such that the Thomas-Fermi screening turns on, so that $|\sqrt{\epsilon(\omega)}|\sim{\cal O}(1)$ in the limit of $\omega R\ll1$. At low frequency $\omega\lesssim0.1\,$eV, the emissivity coefficient $E(\omega)$ cannot be approximated in the limit of a large refractive index $|\sqrt{\epsilon(\omega)}|\gg1$ as assumed in  Eq. \eqref{eq:E omega approx}. Because of this finite-size effect, one should expand $E(\omega)$ in the limit of $|\sqrt{\epsilon(\omega)}|\ll(\omega R)^{-1}$ (see e.g. \cite{bohren:1998absorption}):
\begin{equation}
	\label{eq:E omega screen}
	E(\omega)
	\approx4\omega R\,{\rm Im}
	\left\{\frac{\epsilon(\omega)-1}{\epsilon(\omega)+2}\right\}
	\sim (\omega R)\cdot E^{\rm (FS)}(\omega)\,\qquad
	(|\sqrt{\epsilon(\omega)}|\,\omega R\ll1)\,
\end{equation} 
where $E^{\rm (FS)}(\omega)$ is the emissivity coefficient estimated by FS [Eq. \eqref{eq:E omega approx}]. The suppression factor \( \omega R \) is significant for low-frequency emission \( \omega \ll 0.1 \, \text{eV} \). By neglecting this suppression factor, \cite{Flambaum:2021xub,Flambaum:2021awu,Flambaum:2022wcs} make erroneous predictions regarding the AQN emission in the radio frequency range. For example, FZ proposed in \cite{Forbes:2008uf} that AQNs may explain the observed excess of microwave emission \( \omega \sim 10^{-4} \, \text{eV} \) detected by WMAP (also known as the "WMAP haze") through thermal emission from non-relativistic bremsstrahlung positrons in the electrosphere. FS argued in \cite{Flambaum:2021awu} that this was incompatible with their estimation if thermal emission is dominated by oscillations of the charge density in the electrosphere, rather than by bremsstrahlung radiation. However, charge oscillations are negligible in the low-frequency emission, with a suppression factor of \( \omega R \sim 10^{-3} \) in the case of the WMAP haze, even if we assume their estimate is justified (which is not the case, since the low-frequency limit cannot be properly treated using the traditional "Drude metal" model). It is precisely this frequency range that plays a crucial role in \cite{Forbes:2008uf} in resolving the WMAP haze puzzle.

It is also questionable whether an AQN can be considered a traditional "Drude metal". In condensed matter physics, metals consist of ions embedded in a sea of free electrons, with (either continuous or discrete) translational invariance. This assumption is central to the Drude model and its generalized form, the Lindhard theory. The Lindhard dielectric function \eqref{eq:epsilon q omega} is derived from a perturbative expansion of quantum plane waves. However, for an AQN, the Lindhard dielectric function (and consequently, the Drude dielectric function) must be fundamentally different because the perturbation is based on the expansion of spherical waves rather than plane waves. This difference arises because the chemical potential \( \mu \) and density vary strongly as functions of the radius (distance from the core). Significant modifications occur on scales of the order of \( 10^{-8} \, \text{cm} \), which were entirely overlooked in FS’s calculations. In contrast, in conventional metals, the chemical potential is assumed to be constant over a large sample. Therefore, an AQN is not a traditional metal in the sense of standard condensed matter physics. It does not exhibit translational symmetry nor a uniform ion-electron sea. Instead, an AQN has spherical symmetry and a distinct separation between its quark core and the electrosphere, which is influenced by QCD substructure from the axion domain wall. Additionally, the electrosphere is not uniform but undergoes a steep variation in electron (or positron) density along the radial direction \cite{Forbes:2008uf,Forbes:2009wg}. As a result, an AQN is better modeled as a large, single atom-like object with central symmetry, rather than as a metal.

\subsection{Summary}

We conclude this appendix with the following general remark. We do not claim that the spectral density derived in \cite{Forbes:2008uf,Forbes:2009wg} is exact. It is clear that some assumptions, such as the mean field approximation used by FZ, cannot be valid across the entire parameter space \footnote{In particular, for the thermal emission of an AQN in a very  high density/temperature regime
  there are many additional effects which had not been properly accounted for in the original works \cite{Forbes:2008uf,Forbes:2009wg}. This is because 
  many  positrons may leave the electro-sphere making AQN to become an object carrying  very large negative electric charge. It obviously requires   modification of  the mean field computations by imposing a different boundary condition accounting for this long range interactions due to the charge. This effect will obviously have   many    important observational consequences, see e.g. recent applications for the AQN propagating in dense environment, and we refer to recent papers \cite{Ge:2020cho,Liang:2021rnv,Liang:2021wjx} for the details.}. Nevertheless, we believe that the original FZ approach to the spectrum provides a better description of the complex physics of emission from the AQN's surface, as it is based on solid theoretical foundations. Therefore, we retain the original formulas from \cite{Forbes:2008uf,Forbes:2009wg} in the main body of this work.

\section{\editt{Comparison of emission spectra models}}
\label{app:compare-models}
% {\color{red} @Michael, Appendix B is edited now. You should find a good title for this appendix, and improve the text below. 

% To reference this appendix in the main text, I may suggest putting it somewhere in section \ref{sec:results}, rather than section \ref{subsec:aqn-emission}. This is because at the end of this appendix, we have an estimate of the AQN mass range ($m_{\rm AQN}\gtrsim5\,$kg) predicted based on the FS spectrum. This is relevant to the result in section \ref{sec:results}, where we do claim [100-500] g is preferred values according to our spectrum. You may either reference this appendix using one short sentence following the claim of [100-500] g, or use a footnote, or whatever method you feel reasonable}

% {\color{red} @Xunyu thank you for making these improvements. Why is there a factor of 16 in (B.1) but a factor of 8 in (2.3)? @Michael, it is because (B.1) is the integral $\int \rmd\nu F_\nu$, while (2.3) is $F_\nu$}

\editt{In this appendix, we show that incorrectly using the FS spectrum can lead to a misleading interpretation in this work. As discussed in Appendix \ref{app:emission}, the FS spectrum has many potential pitfalls in its derivation. We use equation \eqref{eqs:F_nu etc.} in this work.} \editt{Specifically,} the total thermal emission used in this work \editt{is} \cite{Majidi:2024-aqn-glow-pt-1}:
\begin{equation}
F
\equiv\int\rmd\nu F_\nu
\approx\frac{16\alpha^{5/2}k_B^4}{3\pi\hbar^3c^2}T_{\rm AQN}^4
\left(\frac{k_BT_{\rm AQN}}{m_ec^2}\right)^{1/4}\,.
\end{equation}
\editt{In comparison, }the FS spectrum in the low energy limit [$k_B T_{\rm AQN}^{\rm(FS)}<50{\rm\,eV}$] \editt{can be found in Ref.} \cite{Flambaum:2021awu}:
\begin{equation}
\begin{aligned}
F^{\rm(FS)}
&\approx0.87\times10^{-4}\cdot
\frac{\pi^2}{60c^2\hbar^3}
\left[k_BT_{\rm AQN}^{\rm(FS)}\right]^4
\sqrt{\frac{k_B T_{\rm AQN}^{\rm(FS)}}{{1\rm\,eV}}}\,,  \\
%&=F_{\rm tot}\times49.5\left(\frac{T_{\rm AQN}^{\rm(FS)}}{T_{\rm AQN}}\right)^{17/4}
%\times\left(\frac{T_{\rm AQN}^{\rm(FS)}}{1\rm\,eV}\right)^{1/4}
\end{aligned}
\end{equation}
where $T_{\rm AQN}^{\rm(FS)}$ is only a parameter, not the internal temperature of the AQN. Since the total annihilation energy \editt{per annihilation cross section is the same, we require}
\begin{equation}
\editt{\frac{F^{\rm(FS)}(T_{\rm AQN}^{\rm(FS)})}{\sigma_{\rm ann}(T_{\rm AQN}^{\rm(FS)})}
=\frac{F(T_{\rm AQN})}{\sigma_{\rm ann}(T_{\rm AQN})}\,,}
\end{equation}
\editt{where the annihilation cross section $\sigma_{\rm ann}$ is determined by the AQN temperature \cite{Majidi:2024-aqn-glow-pt-1}:}
\begin{equation}
\editt{\sigma_{\rm ann}(T_{\rm AQN})
=\frac{8\pi\alpha c^2}{\hbar^2}m_e^2 R^4
\left(\frac{T_{\rm AQN}}{T_{\rm gas,eff}}\right)^2
\sqrt{\frac{k_B T_{\rm AQN}}{m_e c^2}}\,.}
\end{equation}
\editt{It implies:}
\begin{equation}
\editt{T_{\rm AQN}^{\rm(FS)}
\approx \frac{1}{8} T_{\rm AQN}\,,}
\end{equation}
where we assume $k_BT_{\rm AQN}\sim(0.1-10){\rm\,eV}$, the FUV energy in this work. \editt{From Eqs. \eqref{eq:epsilon_nu} and \eqref{eq:Phi_discretized}, we estimate the flux as:}
\begin{equation}
\frac{\Phi_\lambda^{\rm(FS)}}{\Phi_\lambda}
\approx\left.\frac{F_\nu^{\rm(FS)}(T_{\rm AQN}^{\rm(FS)})}{F_\nu(T_{\rm AQN})}
\right|_{T_{\rm AQN}^{\rm(FS)}
= \frac{1}{8} T_{\rm AQN}}\,,\qquad
F_\nu^{\rm(FS)}
\equiv2\pi\frac{\rmd}{\rmd\omega}F_\omega^{\rm(FS)}\,.
\end{equation}
If using the FS spectrum, one finds \editt{$\Phi_\lambda^{\rm(FS)}\lesssim0.02\Phi_\lambda$ near $\lambda=1550$\,\AA. It would correspond to a completely different AQN mass range $m_{\rm AQN}\gtrsim5{\rm\,kg}$, well outside the allowed mass window. The source of this problem is a number of incorrect assumptions used in the FS spectrum as discussed in Appendix \ref{app:emission}. }

\bibliographystyle{JHEP}
\bibliography{JCAP_aqn_GALEX_uv.bib}

\end{document}